\documentclass[aps,pre,twocolumn,amsfonts,amsmath,showpacs,showkeys,floatfix,reprint]{revtex4-1}

\usepackage{amsmath,graphicx,amssymb,amsfonts}
\usepackage{natbib}

\begin{document}

\title{Linked by dynamics: \\wavelet--based mutual information rate
as a connectivity measure\\
 and scale-specific networks}


\author{Milan Palu\v{s}}
\email[]{mp@cs.cas.cz}
\homepage[]{http://www.cs.cas.cz/mp/}
\affiliation{Department of Nonlinear Dynamics and Complex
Systems,
Institute of Computer Science, Academy of Sciences of the
Czech Republic,
 Pod vod\'arenskou v\v{e}\v{z}\'{\i}~2, 182~07~Prague~8,
Czech~Republic}


\date{\today}

\begin{abstract}
Experimentally observed networks of interacting dynamical systems
are inferred from recorded multivariate time series by evaluating
a statistical measure of dependence, usually the cross-correlation
coefficient, or mutual information. These measures reflect
dependence in static probability distributions, generated by
systems' evolution, rather than coherence of systems' dynamics.
Moreover, these ``static'' measures of dependence can be biased
due to properties of dynamics underlying the analyzed time series.
Consequently, properties of local dynamics can be misinterpreted
as properties of connectivity or long-range interactions. We
propose the mutual information rate as a measure reflecting
coherence or synchronization of dynamics of two systems and not
suffering by the bias typical for the ``static'' measures. We
demonstrate that a computationally accessible estimation method,
derived for Gaussian processes and adapted by using the wavelet
transform, can be effective for nonlinear, nonstationary and
multiscale processes.
 The discussed problem and the proposed method are illustrated using
 numerically generated data of coupled dynamical systems as well as gridded reanalysis
 data of surface air temperature as the source for the construction of climate networks.
 In particular, scale-specific climate networks are introduced.\\

 To appear in: {\bf Tsonis, A.A. (ed.): Advances in Nonlinear
 Geosciences, Springer, 2017.}
\end{abstract}





\maketitle

\section{Introduction}

``More is different,'' the simple sentence of the most creative
\cite{creat} physicist P.W. Anderson \cite{more} reflects the
complex reality in which the behavior of complex systems,
consisting of many interacting elements, cannot be explained by a
simple extrapolation of the laws describing the behavior of a few
elements. Studying systems of many interacting elements as complex
networks
\cite{RevModPhys.74.47,boccaPR,newman2006structure,havlin12} is an
intensively developing paradigm in which statistical physics
embraced the graph theory. In the graph-theoretical
characterization of complex networks, a network is considered as a
 graph $G=(V,E)$, where $V$ is a set of
nodes (or vertices) and $E$ is a set of edges (or links) where
each edge represents a connection between two nodes. In the case
of weighted graphs a weight $w_{i,j}$ is assigned to each edge
$e_{i,j}$, connecting the vertices $v_i$ and $v_j$, by the weight
function $W:E \rightarrow \mathbb{R}$. The graph $G=(V,E)$ is
characterized by the adjacency matrix $A$ whose elements $a_{i,j}
= w_{i,j}$; and $a_{i,i} = 0$ by definition. We will consider
undirected graphs, i.e. $a_{i,j} = a_{j,i}$. A special case of
graphs are unweighted graphs, also known as binary graphs, since
$a_{i,j}$ can attain either the value 1 if $e_{i,j} \in E$, or the
value $0$ otherwise.

In this study we will consider networks of interacting, possibly
stochastic, dynamical systems. In the network paradigm, each
system represents a node of the network. Consider that the
interactions among the nodes (dynamical systems) are not known.
However, we can observe and record evolution of each dynamical
system. A series of measurements done on such a system in
consecutive instants of time $t = 1, 2, \dots$ is usually called a
time series $\{x(t)\}$. In order to infer a network from a
multivariate time series $\{x_i(t)\}$ usually some measure of
statistical dependence between components $\{x_i(t)\}$ and
$\{x_j(t)\}$, recorded from the nodes $v_i$ and $v_j$,
respectively, is estimated. This measure, or a transformation
thereof, is considered as a weight $w_{i,j}$ assigned to the edge
$e_{i,j}$. The networks of this type are known as interaction
networks~\cite{Bialonski2010} or functional networks. The latter
term have been spread from neurophysiology where the statistical
association of neural activities in two distinct parts of the
brain is called the functional connectivity \cite{friston94}, as
opposed to a structural, anatomical connectivity given by an
existence of a physical link \cite{Bullmore2009}.
Neurophysiology is probably the most active and influential
scientific field where the functional networks are constructed and
studied; making use of a huge amount of multivariate data
recording various modes of brain activity
\cite{Bullmore2009,achard06,stam07}. The interaction networks,
however, are studied also in different areas such as climatology
\cite{tsonis2004architecture,tsonis2006,yamaninoprl,yamasaki09,donges2009,donges2009backbone,gangclidym}
or economy and finance \cite{economicnet,correlfinnet}.
Since the existence of a link in an interaction network is
inferred from an estimate of a statistical dependence measure,
the strength and even the existence of a link bear some level of
uncertainty.
Kramer et al.~\cite{kramerPRE} propose a systematic statistical
procedure for the inference of functional connectivity networks
from multivariate time series yielding as the output both the
inferred network and a quantification of uncertainty of the number
of edges. Palu\v{s} et al.~\cite{npgnet11} present differences in
the topology of interaction networks with edges derived either
from the largest absolute correlations or from the statistically
most significant absolute correlations. Bialonski et
al.~\cite{Bialonski2010} demonstrate that a spatial sampling can
lead to an occurrence of spurious structures in interaction
networks constructed from time series sampled in spatially
extended systems and propose tailored random networks as a
suitable null hypothesis to be tested \cite{Bialonski2011}. Hlinka
et al.~\cite{hlinkaswn} observed that a spurious small-world
topology emerged in interaction networks constructed using
correlations of time series generated by randomly connected
dynamical systems. While Bialonski et al.~\cite{Bialonski2010}
attribute spurious topologies to sampling problems and
finite-precision, finite-length time series, Hlinka et
al.~\cite{hlinkaswn} see the problem in partial transitivity -- an
inherent property of the correlation coefficient. Also Zalesky et
al.~\cite{Zalesky20122096} observed that the networks in which
connectivity was measured using the correlation coefficient were
inherently more clustered than random networks, while partial
correlation networks were inherently less clustered than random
networks. Therefore, in a similar line with Bialonski et
al.~\cite{Bialonski2011}, also Zalesky et
al.~\cite{Zalesky20122096} propose to use a sort of null networks
in order to explicitly normalize for the inherent topological
structure found in the correlation networks.

In this study we will focus on the dynamics underlying time series
used for the construction of interaction networks. We will
demonstrate how ``dynamical memory'' influences the bias in
estimations of ``static'' dependence measures such as the absolute
correlation coefficient or the mutual information. We will propose
the mutual information rate as a measure reflecting dependence of
dynamics of two systems or processes. We will introduce a
computationally accessible algorithm that can be effective for
quantification of the coherence or synchronization of nonlinear,
nonstationary and multiscale processes and thus can be used for
the construction of interaction networks from experimental time
series recorded in natural complex systems.

\section{Dependence}\label{dep1}

Consider two discrete random variables $X$ and $Y$  with sets of
values $\Xi$ and $\Upsilon$, respectively. The probability
distribution function (PDF) $p_{X}(x) $ for the variable $X$,
 for simplicity denoted as $p(x) $,
 is  $p(x) =$ Pr$\{X=x\}$, $x \in \Xi$. The probability distribution function
 $p(y)$ for the variable $Y$ is defined in the full analogy; and the
 joint PDF $p(x,y)$ is Pr$\{(X,Y)=(x,y)\}$, $x \in \Xi$, $y \in
 \Upsilon$. Uncertainty in a random variable, say $X$, is
 characterized by its entropy
\begin{equation}\label{hx}
H(X) = - \sum _{x \in \Xi} p(x) \log p(x).
\end{equation}

The joint entropy $H(X,Y)$ of $X$ and $Y$ is
\begin{equation}\label{hxy}
H(X,Y) = - \sum _{x \in \Xi} \sum _{y \in \Upsilon} p(x,y) \log
p(x,y).
\end{equation}

The two variables $X$ and $Y$ are independent if and only if
$p(x,y)=p(x)p(y)$, i.e.
$$
\log \frac{p(x,y)}{p(x)p(y)} = 0.
$$
 The average digression from independence, i.e., the
   averaged value of $\log \frac{p(x,y)}{p(x)p(y)}$ is known as
mutual information
\begin{equation}\label{Ixy}
 I(X;Y) =    \sum _{x \in \Xi} \sum _{y \in \Upsilon}
p(x,y) \log \frac{p(x,y)}{p(x)p(y)}.
\end{equation}

 The mutual information can be expressed using the
entropies (\ref{hx}), (\ref{hxy}) as
\begin{equation}\label{Ihxhy}
I(X;Y) = H(X) + H(Y) - H(X,Y).
\end{equation}
Thus the mutual information $I(X;Y)$ quantifies the decrease of
uncertainty in $H(X,Y)$ due to the dependence between $X$ and $Y$,
i.e., it measures the average amount of common information,
contained in the variables $X$ and $Y$. The mutual information is
a measure of general statistical dependence for which the
following statements hold:
\begin{itemize}
\item $I(X;Y) \ge 0$,

\item $I(X;Y) = 0$ iff $X$ and $Y$ are independent.

 \end{itemize}

 In practice, however, the PDF's are not known and we only have
 a set of measurements $\{x_1, x_2, \dots , x_N\}$ for the variable $X$
 and  $\{y_1, y_2, \dots , y_N \}$ for the variable $Y$.
 Estimation of the entropies (\ref{hx}), (\ref{hxy}) and the mutual
 information (\ref{Ixy}) can be done using some of suitable
 estimators, for review see Ref.~\cite{katkaPR}.

 A common measure of
 linear dependence is the (Pearson's) correlation coefficient. First,
 we compute the mean of all measurements $\{x_1, x_2, \dots , x_N\}$
 as
$$ \bar x = \frac{1}{N} \sum_{i=1}^{N} x_i$$
and the variance $$\sigma ^2 = \frac{1}{N-1} \sum_{i=1}^{N} (x_i -
\bar x)^2 $$ and transform the measurements into a data with a
zero mean and a unit variance
\begin{equation}\label{normx}
\widetilde{x_i} = \frac{x_i - \bar x}{\sigma}.
\end{equation}
After the same procedure with the measurements of the variable
$Y$, the correlation coefficient of $X$ and $Y$ is
\begin{equation}\label{cxy}
C(X,Y) = \frac{1}{N} \sum_{i=1}^{N} \widetilde x_i \widetilde y_i
.
\end{equation}
Without loss of generality, in the following we will suppose that
considered data or time series have (or have been transformed in
order to have) a zero mean and a unit variance.

Suppose that the variables $X$ and $Y$ have a bivariate Gaussian
distribution. Then their mutual information $I(X;Y)$ can be
expressed using their correlation coefficient $C(X,Y)$ (see, e.g.
Ref.~\cite{rd0} and references therein)
\begin{equation}\label{IG}
I(X;Y) = -\frac{1}{2} \log \big( 1 - C^2(X,Y) \big).
\end{equation}

The correlation coefficient (\ref{cxy}) and the mutual information
(\ref{Ixy}) are the measures of dependence which reflect the
digression of the ``static'' bivariate distribution $p(x,y)$ from
the product $p(x)p(y)$. We use the term ``static'' in order to
stress that both the correlation coefficient (\ref{cxy}) and the
bivariate PDF $p(x,y)$ which determines the mutual information
(\ref{Ixy}) are given by the set of pairs $\{(x_1,y_1), (x_2,y_2),
\dots , (x_N,y_N)\}$ irrespectively of the order of the pairs. Any
permutation of the pairs $(x_i,y_i)$ yields the same result.

\section{Dynamics}\label{dyn1}

Let us consider $n$ discrete random variables $X_1, \dots , X_n$
with values $(x_{1}, \dots , x_{n}) \in \Xi_{1} \times
\dots \times \Xi_{n}$. The PDF
 for  an individual $X_i$ is  $p(x_i) = $
Pr$\{X_i=x_i\}$, $x_i \in \Xi_i$, the joint PDF  for the $n$
variables  $X_1, \dots , X_n$ is  $p(x_{1}, \dots , x_{n}) = $
Pr$\{(X_{1}, \dots , X_{n}) = (x_{1}, \dots , x_{n})\}$. The joint
entropy of the $n$ variables $X_1$,$\dots$, $X_n$ with the joint
PDF $p(x_1 , \dots , x_n)$ is
\begin{displaymath}
H(X_{1}, \dots, X_{n}) =
\end{displaymath}
\begin{equation}\label{hxn}
- \sum _{x_{1} \in \Xi_{1}} \dots \sum _{x_{n} \in \Xi_{n}}
p(x_{1}, \dots , x_{n}) \log p(x_{1}, \dots , x_{n}).
\end{equation}

A stochastic process $\{X_{i}\}$ is an indexed sequence of random
variables $X_{1}, \dots ,X_{n}$, characterized by the joint PDF
$p(x_{1}, \dots , x_{n})$. Uncertainty in a variable $X_i$ is
characterized by its entropy $H(X_i)$. The rate at which a
stochastic process ``produces'' uncertainty is measured by its
entropy rate
\begin{equation}\label{hrate}
h = \lim _{n \rightarrow \infty} \frac{1}{n} H(X_{1}, \dots
,X_{n}).
\end{equation}

In practice we will deal with a  time series $\{x(t)\}$, $t = 1,
2, \dots, N$. While considering measurements $\{x_1, x_2, \dots ,
x_N\}$ of a random variable $X$, its values $x_i$ are typically
considered mutually independent, i.e., obtained by independent,
random draws from a PDF $p(x)$. On the other hand, a  time series
$\{x(t)\}$ reflects a temporal evolution of a process or a system,
and typically the values $x(t)$ and $x(t+\tau)$, where $\tau$ is a
time lag, are not independent. The level of dependence between
$x(t)$ and $x(t+\tau)$ reflects a ``dynamical memory'' of the
temporal evolution of an underlying process or system. The
decrease of the dependence between $x(t)$ and $x(t+\tau)$, with
increasing $\tau$, i.e., the rate at which a process ``forgets''
its history depends on complexity of the temporal evolution of a
process or a system and we will refer to this complexity as
``temporal dynamics,'' or shortly as ``dynamics.''

Since a time series $\{x(t)\}$ reflects the dynamics of an
underlying process or system, a stochastic process $\{X_{i}\}$
characterized by the joint PDF $p(x_{1}, x_2 \dots , x_{n})$ which
typically differs from the product $p(x_{1})p(x_2) \dots
p(x_{n})$, is an appropriate theoretical concept for the study of
time series. Thus a time series is considered as a realization of
a stochastic process $\{X_{i}\}$, and should not be equated with a
set of measurements of a single variable $X$ with a PDF $p(x)$.
The entropy rate (\ref{hrate}) is a useful characterization of the
dynamics of a system or a process underlying the time series
$\{x(t)\}$. In information theory the entropy rate (\ref{hrate})
is considered as a measure of production of information of an
information source \cite{thomas}.

Alternatively, a time series $\{x(t)\}$ can be considered as a
projection of a trajectory of a dynamical system, evolving in a
measurable state space. A.~N.~Kolmogorov, who introduced the
theoretical concept  of classification of dynamical systems by
information rates, was inspired by information theory and
generalized the notion of the entropy of an information source.
The Kolmogorov-Sinai entropy (KSE thereafter) or metric entropy
\cite{petersen}
 is a topological invariant, suitable for the
classification of dynamical systems or their states, and is
related to the sum of the system's positive Lyapunov exponents
\cite{pesin}. The concept of entropy rates is common to theories
based on philosophically opposite  assumptions (randomness vs.
determinism) and is ideally applicable for the characterization of
complex processes, where possibly deterministic rules are always
accompanied by random influences.

As a potentially useful quantitative characterization of the
dynamics, the entropy rate has become a target of many numerical
algorithms using experimental time series as their input.
Particularly intensive development, focused on the estimation of
the metric entropy, has started with the advent of the methods for
the reconstruction of chaotic dynamics in the 1980's. Grassberger
and Procaccia \cite{grassberger1983KSE} used the concept of
R\'enyi entropy \cite{thomas} to redefine the KSE in the terms of
the R\'enyi entropy of order two and proposed an estimator of the
metric entropy $K_2$ using their celebrated correlation integral
\cite{GrassbergerPhysD}. The method has been extended into
numerous version, e.g. by Cohen and Procaccia \cite{cohen85}.
Schouten et al.~\cite{PhysRevE.49.126} treated the correlation
integral as a probability distribution and derived a
maximum-likelihood estimator of the KSE.
 Pawelzik and Schuster \cite{PhysRevA.35.481} consider the
full spectrum of generalized metric entropies $K_q$.
Fraser \cite{fraser1989information} pointed to an interesting
relation between an n-dimensional version of the mutual
information and the KSE of a dynamical system underlying studied
time series. Palu\v{s} \cite{palus1997kolmogorov} studied this
relation in detail and confirmed its validity by comparing the KSE
estimates with the values of the positive Lyapunov exponents of
the studied chaotic systems. Reliable KSE estimates, however,
require large amounts of data. Therefore Palu\v{s} \cite{cer}
proposed ``coarse-grained entropy rates'' which relate the KSE to
the rate of the decrease of a finite-precision mutual information
of a time series and its time-lagged twins. Also bounded by a
finite precision and a limited amount of real data, Pincus
\cite{pincusPNAS} introduced an approximate entropy based on a
difference of the correlation integrals.

The entropy rate reflects how quickly a system ``forgets'' its
history. In the case of chaotic dynamical systems the metric
entropy is related to a time interval which a dynamical system
takes to return to a close vicinity of some of its previous
states. Baptista et al.~\cite{Baptista20101135} propose two
formulas to estimate the KSE and its lower bound  from the
recurrence times of chaotic systems. The recurrence plots
\cite{Marwan2007237} give a number of useful dynamical quantities
including the KSE.

A time series of measurements of a finite precision can be
conveniently converted into a sequence of symbols from a finite
set of values. Bandt and Pompe \cite{bandt2002permutation}
introduced the concept of permutation entropy for symbolic
sequences and demonstrate its relations to the KSE. Lesne et
al.~\cite{lesne09} studied entropy rate estimators for short
symbolic sequences based on block entropies and Lempel-Ziv
complexity \cite{lempelziv}. 
Kennel et al.~\cite{kennel2005} developed an algorithm for
estimating the entropy rate of Markov models using weighted
context trees.
The entropy rates can also be computed using the causal state
machine based estimator
\cite{crutch-youngPRL,shalcrutchcompmech,haslinger2010shalizi}.

Let us return from symbolic sequences to continuous stochastic
processes. Let a stochastic process $\{X_{i}\}$
 is a zero-mean,
stationary, Gaussian process with power spectral density
$\Phi(\omega)$, where $\omega$ is a normalized frequency. Then its
entropy rate $h_G$, apart from a constant term, is
\cite{pinsker,gser}
\begin{equation}\label{hg}
h_G = \frac{1}{2\pi} \int^{2\pi}_{0} \log \Phi(\omega) d \omega.
\end{equation}

\section{Dynamics and connectivity}\label{dyco}

In order to understand the notion of temporal dynamics of a
process and its characterization using the entropy rate,  let us
consider the autoregressive process (ARP)
\begin{equation}\label{arp}
x(t) =  c \sum_{k=1}^{10} a_k x(t-k) + \sigma e(t),
\end{equation}
where $a_{k=1,..,10} = {0, 0, 0, 0, 0, .19, .2, .2, .2, .2}$,
$\sigma = 0.01$ and $e(t)$ is a Gaussian noise with a zero mean
and a unit variance. The parameter $c$ modulates the proportion of
the deterministic part of the process which is a function of the
history of the process, to the noise part of the process. The
greater the coefficient $c$, the stronger the memory, i.e., the
dependence between $x(t)$ and $x(t+\tau)$.
\begin{figure}
 \includegraphics[width=1.0\columnwidth]{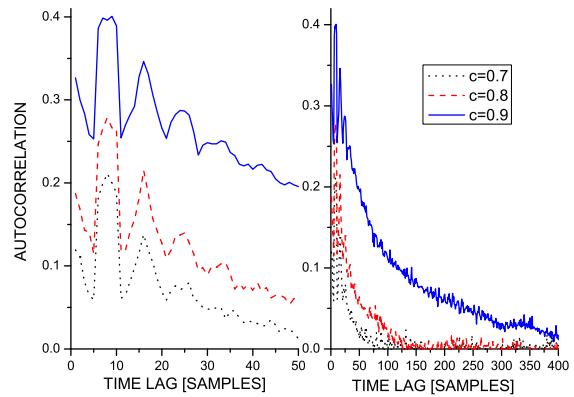}
 \caption{(Color online) Autocorrelation function for the
 autoregressive process (\ref{arp}) for different values of the
 coefficient $c$: $c=0.7$ (the dotted black line), $c=0.8$ (the dashed red line),
 and $c=0.9$ (the solid blue line). Time lags 1--400 samples
 (right panel), the detail for time lags 1--50 samples (left
 panel).
 }\label{aracf}
 \end{figure}
This effect is demonstrated in Fig.~\ref{aracf}, where the
autocorrelation function $C\big(x(t),x(t+\tau)\big)$ as a function
of the time lag $\tau$ is plotted for different values of the
coefficient $c$. For $c=0.7$ (the dotted black line) the
autocorrelation function (ACF) has the lowest values and vanishes
(fluctuates with values close to zero) for time lags around 100
samples; for $c=0.8$ (the dashed red line) the ACF has higher
values and vanishes about the time lag equal to 150 samples, while
for $c=0.9$ (the solid blue line) the ACF has the largest values
and requires more than 400 samples of the time lag to vanish. The
ACF reflects the fact that increasing $c$ the dynamical memory of
the process (\ref{arp}) is stronger and longer lasting.
\begin{figure}
 \includegraphics[width=1.0\columnwidth]{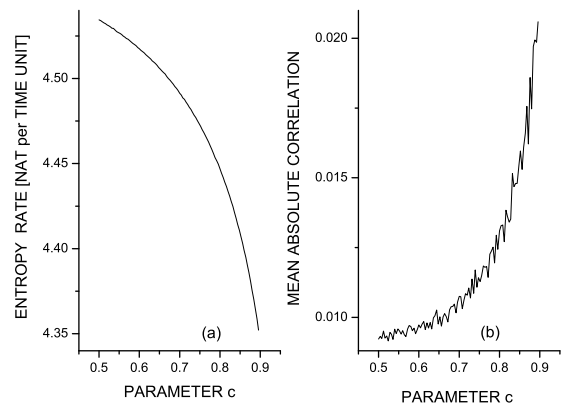}
 \caption{(a) Entropy rate $h_G$ for the
 autoregressive process (\ref{arp})
as a function of the parameter $c$. (b) Dependence of the mean
absolute cross-correlation between independent realizations of the
 autoregressive process (\ref{arp}) on the
 parameter $c$.
 }\label{arratemac}
 \end{figure}

How these differences in the dynamical memory, or in the dynamics
are reflected in the entropy rate? We generate realizations of the
ARP (\ref{arp}) with different $c$ and compute the entropy rates
$h_G$ according to Eq.~(\ref{hg}). Figure~\ref{arratemac}a
presents the entropy rate $h_G$ for 100 realizations of the ARP
(\ref{arp}) with $c$ increasing from 0.5 to 0.9. The entropy rate
of such ARP's monotonically decreases with increasing $c$. A
higher entropy rate means that the process generates uncertainty
at a higher rate so that it forgets its history more quickly.
Predictability of a process with a higher entropy rate is worse
and possible for a shorter prediction horizon than predictability
of a process with a lower entropy rate.

Time series $\{x_i(t)\}$ recording temporal evolution of different
systems or subsystems of a complex system might reflect different
dynamics yielding different entropy rates. As we have noted in the
Introduction, the connectivity in complex networks constructed
from multivariate time series, i.e., the existence and the
strength of links between nodes are inferred using dependence
measures such as the mutual information (\ref{Ixy}) and the
correlation (\ref{cxy}). The absolute value of the latter is
typically used, while the mutual information is always
non-negative. Applying the definitions (\ref{Ixy}) and (\ref{cxy})
to time series $\{x(t)\}$ and $\{y(t)\}$, they are treated as sets
$\{x_i\}$ and $\{y_i\}$ of measurements of random variables $X$
and $Y$. The computed $C(X,Y)$ or $I(X;Y)$ do not reflect the
dynamics of $\{x(t)\}$ and $\{y(t)\}$. Indeed, the pairs
$(x(t),y(t))$ would yield the same values of $C(X,Y)$ or $I(X;Y)$
independently of their temporal order.
The computed values of $C(X,Y)$ or $I(X;Y)$ are, however, only
estimates of the true dependence between  processes generating the
datasets  $\{x(t)\}$ and $\{y(t)\}$. The estimates have some bias,
giving a mean digression from the true value, and a variance
giving the range of fluctuations of the estimates around their
mean value.

\begin{figure}
 \includegraphics[width=1.0\columnwidth]{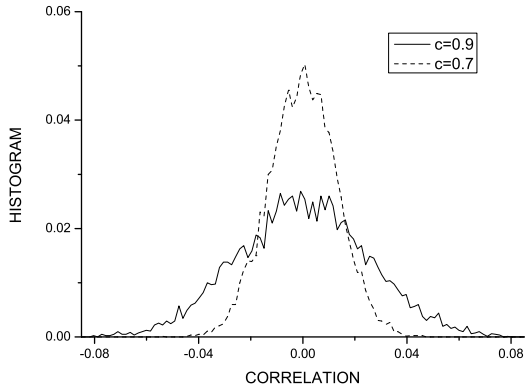}
 \caption{Histograms of cross-correlations between independent realizations
of the
 autoregressive process (\ref{arp}) for two different values of
the parameter $c$.
 }\label{corhist}
 \end{figure}
Using the above defined ARP (\ref{arp}) we can study the behavior
of the correlation estimates for time series with different
dynamics.
  In particular, we can
generate realizations of the ARP (\ref{arp}) with different $c$'s
and thus with different entropy rates.  Now, let us study the
distribution of the cross-correlations between {\it independent}
realizations of the process (\ref{arp}) for different values of
the parameter $c$. For each $c$ we generate 8192 process
realizations, each realization consisting of 16,384 samples.
Figure~\ref{corhist} presents histograms of cross-correlations
between independent realizations of ARP (\ref{arp}) for two
different $c$'s. The mean value is always correctly equal to zero,
however, the variance increases with increasing $c$, i.e., with
decreasing the entropy rate. As a consequence, when considering
the {\it absolute} correlations, or a non-negative dependence
measure such as the mutual information, its mean value has an
increasing upward bias with the decreasing entropy rate. This
effect is illustrated in Fig.~\ref{arratemac}b. In this example
the bias in the absolute correlations reaches relatively small
values 0.01 -- 0.02. These values, however, are obtained for time
series of 16,384 samples. In Sec.~\ref{clinet} we will show that
in real time series of 512 samples the bias  can reach such values
as 0.4. For even shorter and/or more regular (lower entropy rate)
time series the bias can be even higher \cite{contempap}.

\section{Mutual information rate}\label{secmir}

Instead of treating  time series $\{x(t)\}$ and $\{y(t)\}$ as sets
$\{x_i\}$ and $\{y_i\}$ of measurements of random variables $X$
and $Y$, now let us consider the time series $\{x(t)\}$ and
$\{y(t)\}$ as realizations of
 stochastic processes $\{X_{i}\}$ and
$\{Y_{i}\}$,
  characterized by PDF's $p(x_{1}, \dots ,
x_{n})$
 and $p(y_{1}, \dots , y_{n})$, respectively. In the analogy of generalization of
 the entropy (\ref{hx}) to the entropy rate (\ref{hrate}) in order
 to characterize dynamics of a process, now we generalize the
 mutual information (\ref{Ixy}) to the mutual information rate
 (MIR)  \cite{thomas} as
\begin{equation}\label{mir}
i(X_{i}; Y_{i}) = \lim _{n \rightarrow \infty} \frac{1}{n}
I(X_{1}, \dots ,X_{n} ; Y_{1}, \dots, Y_{n}).
\end{equation}
While the mutual information $I(X;Y)$ evaluates the difference
between the bivariate PDF $p(x,y)$ and the product of the
univariate PDF's $p(x)p(y)$, the MIR (\ref{mir}) is the limit
value of the mutual information $I(X_{1}, \dots ,X_{n} ; Y_{1},
\dots, Y_{n})$ evaluating the difference between the $2n$-variate
PDF $p(x_{1}, \dots , x_{n}, y_{1}, \dots , y_{n})$ and the
product of the two $n$-variate PDF's $p(x_{1}, \dots ,
x_{n})p(y_{1}, \dots , y_{n})$. The MIR quantifies the dependence
between the sequences of states $X_{1}, \dots ,X_{n}$ of the
process $\{X_{i}\}$ and states $Y_{1}, \dots, Y_{n}$ of the
process $\{Y_{i}\}$. In the case of dynamical systems the MIR
reflects coherent dynamics or a common evolution of two systems
whose trajectories are projected onto the time series $\{x(t)\}$
and $\{y(t)\}$.

For pairs of dynamical systems that are either mixing, or exhibit
fast decay of correlations, or have sensitivity to initial
conditions, Baptista et al.~\cite{baptistaMIR}
 have proposed a way how to calculate MIR and its upper  and lower bounds
  in terms of Lyapunov exponents, expansion rates, and
capacity dimension. In general, estimators of MIR are well
elaborated for symbolic dynamics, extending the estimators of the
entropy rates. Shlens et al.~\cite{shlens2007estimating} further
develop the estimator of Kennel et al.~\cite{kennel2005} and
applied it in order to estimate the information transfer between a
stimulus and neural spike trains. Blanc et al.~\cite{blancMIR}
extended the entropy rate estimator for symbolic sequences
\cite{lesne09} and compared several estimators adapted for the
estimation of the MIR between coupled dynamical systems in a
symbolic representation, including the Lempel-Ziv \cite{lempelziv}
and the causal state machine based estimator
\cite{crutch-youngPRL,shalcrutchcompmech,haslinger2010shalizi}.

Considering continuous stochastic processes, for zero-mean,
Gaussian stochastic processes $\{X_{i}\}$, $\{Y_{i}\}$,
characterized by
 power spectral densities (PSD) $\Phi_{X}(\omega)$, $\Phi_{Y}(\omega)$
 and cross-PSD $\Phi_{X,Y}(\omega)$, the MIR can be expressed
(see Ref. \cite{pinsker}) as

\begin{equation}\label{mirg}
i_G(X_{i}; Y_{i}) = - \frac{1}{4\pi} \int^{2\pi} _0 \log(1 -
|\Gamma_{\small{X,Y}} (\omega)|^2) d\omega,
\end{equation}
using the magnitude-squared coherence
\begin{equation}\label{gamma}
|\Gamma_{X,Y} (\omega)|^2 = \frac{|\Phi_{X,Y}
(\omega)|^2}{\Phi_{X}(\omega)\Phi_{Y}(\omega)}.
\end{equation}

 Now we can return to the ARP (\ref{arp}) and use its independent
 realizations generated with different values of the parameter $c$
 and thus characterized by different entropy rates, in order to
 study the
 bias of dependence measures in relation to dynamics (entropy rate).
\begin{figure}
 \includegraphics[width=1.0\columnwidth]{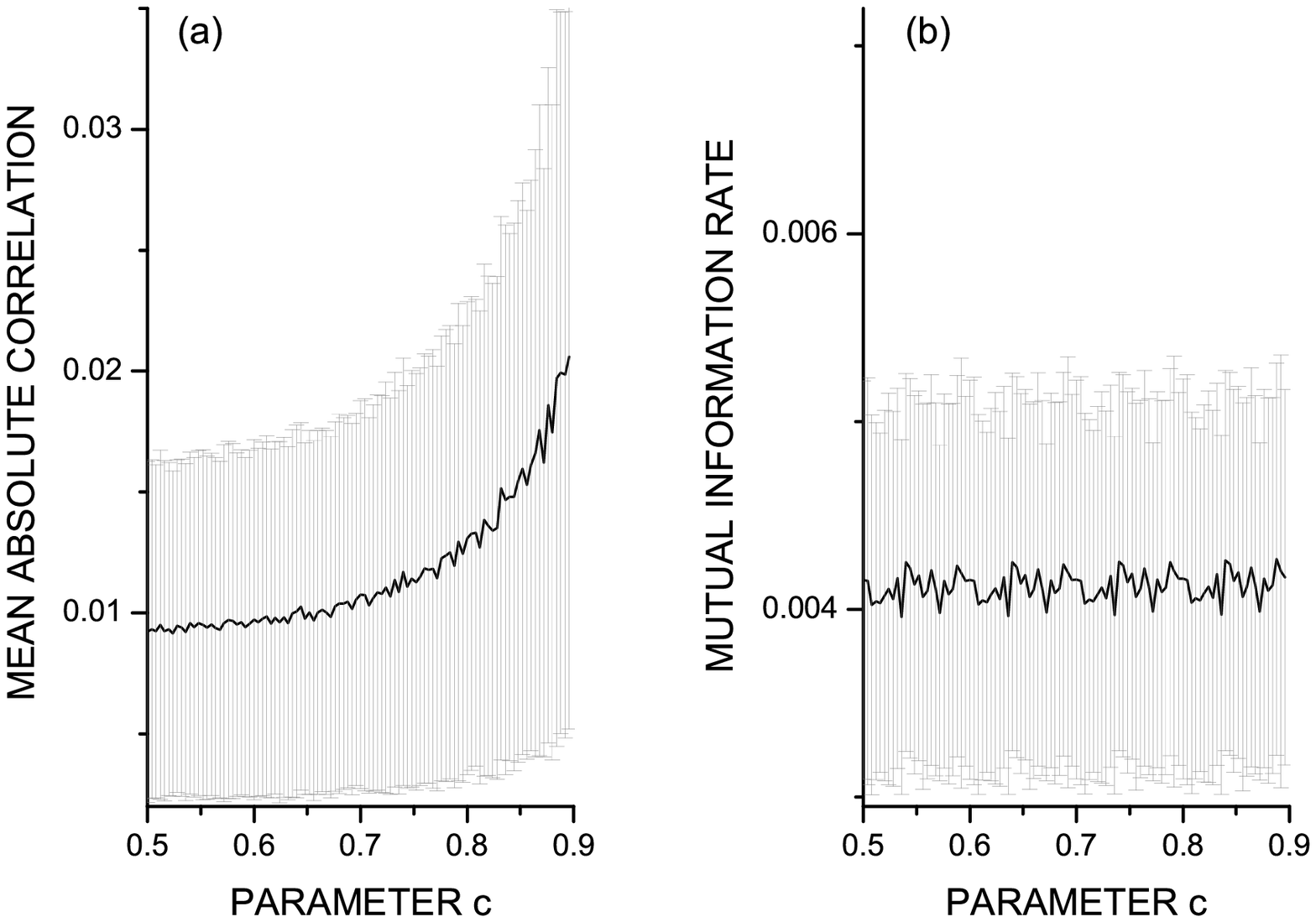}
 \caption{(a) Mean (solid line) and variance (bars $\pm \sigma$ above and below the mean value)  of
the absolute cross-correlation between independent realizations of
the
 autoregressive process (\ref{arp}) as a function of the
 parameter $c$. (b) The same as (a) but for the mutual information rate (\ref{mirg}).
 Note that scales in (a) and (b) are different.
 }\label{cormirsd}

 \end{figure}
In Fig.~\ref{cormirsd}a we study again the absolute
cross-correlations of independent
 realizations the ARP (\ref{arp}) as a function of the parameter $c$. The mean values are
 the same as in Fig.~\ref{arratemac}b, however, here we illustrate also the variance as the
 bars mean$\pm \sigma$. We can see that with increasing $c$ (decreasing the entropy rate) both the
 mean and variance of the absolute cross-correlations increase. Using the computationally feasible
 formula (\ref{mirg}) for the mutual information rate, in Fig.~\ref{cormirsd}b we present
 means and variances for the MIR estimates for independent
 realizations the ARP (\ref{arp}) as a function of the parameter $c$. There is some positive bias,
 represented by the mean MIR, which is low, randomly fluctuating and independent of the
 dynamics of the evaluated time series, i.e., independent of the parameter $c$. Also the variances
 of MIR are independent of $c$, they are practically the same for all values
 of $c$ -- the positions of bars  $\pm \sigma$
 in Fig.~\ref{cormirsd}b are given by the fluctuations in the mean value. The mutual information
 rate is a measure of dependence between dynamics of systems or processes, and unlike the static
 measures, its bias does not depend on the complexity of dynamics.

\section{Information rates of Gaussian processes and dynamical systems}\label{gady}

The formulas (\ref{hg}) for the entropy rate and (\ref{mirg}) for
the mutual information rate of Gaussian processes can be
efficiently evaluated using the fast Fourier transform (FFT). The
question is, however, how applicable are these formulas for
real-world time series recorded from complex, possibly nonlinear
systems. Using a number of paradigmatic chaotic dynamical systems,
Palu\v{s} \cite{gser} inquired a relation between the
Kolmogorov-Sinai entropy of a dynamical system and the entropy
rate of a Gaussian process with the same spectrum as the sample
spectrum of the time series generated by the dynamical system. An
extensive numerical study suggests  that such a relation as a
nonlinear one-to-one function exists when the Kolmogorov-Sinai
entropy varies smoothly with variations of system's parameters,
but is broken near bifurcation points. Although the formula
(\ref{hg}) does not give values numerically close to the true
values of the Kolmogorov-Sinai entropy of studied dynamical
systems, it allows a relative quantification and distinction of
different states of nonlinear systems. In a practical application,
the formula (\ref{hg}) was used in order to characterize changing
complexity of dynamics of neuronal oscillations on route to an
epileptic seizure \cite{jiruska2010}. A strongly nonlinear
character of the neuronal activity of epileptogenic brain regions
has been confirmed, e.g., by Casdagli et
al.~\cite{Casdagli1996381}.

\begin{figure}
 \includegraphics[width=1.0\columnwidth]{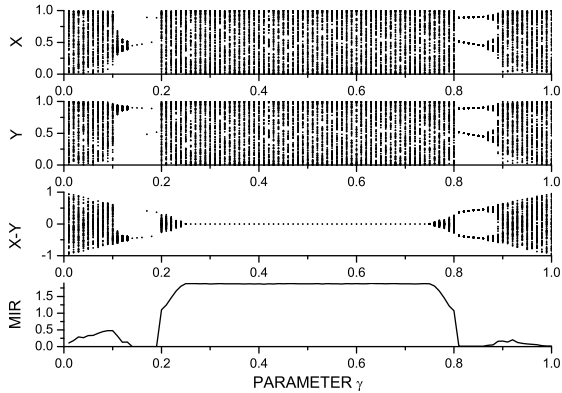}
 \caption{Top three panels: bifurcation diagrams of two
coupled logistic maps (from the top: $x$, $y$, and $x-y$), for the
control parameter value $a = 4$, corresponding to fully chaotic
maps when uncoupled, as a function of the coupling coefficient
$\gamma$. Bottom panel: the mutual information rate (\ref{mirg})
between $\{X\}$ and
 $\{Y\}$, computed using the FFT, as a function of the coupling coefficient
$\gamma$.
 }\label{log2}
 \end{figure}

In order to demonstrate how the formula (\ref{mirg}) for the
mutual information rate of Gaussian processes reflects changes in
the dependence of dynamics of two coupled nonlinear dynamical
systems on their route to synchronization we will use two
well-know dynamical systems with chaotic behavior. As an example
of a discrete-time system let us borrow the symmetrically coupled
logistic maps from Blanc et al.~\cite{blancMIR} where the system
$\{X\}$ is represented by the time series $\{x_n\}$ and
 the system $\{Y\}$ by the time series
 $\{y_n\}$:
\begin{eqnarray}\label{logi2}
\begin{split}
x_{n+1} &= & \gamma f_a(x_{n}) + (1 - \gamma) f_a(y_{n})  \\
y_{n+1} &= & (1-\gamma) f_a(x_{n}) + \gamma f_a(y_{n})
\end{split}
\end{eqnarray}
where $\gamma$ is the coupling coefficient and varies between 0
and 1. The function $f_a(x_{n}) \equiv a x_n (1 - x_{n})$. It is
known that, in the uncoupled case, $a=4$ gives a chaotic behavior.
The latter is demonstrated in the bifurcation diagrams in
Fig.~\ref{log2} where for small $\gamma$ both the system $\{X\}$
and $\{Y\}$ are chaotic and not synchronized.  For $0.13 < \gamma
< 0.2$ a zone of periodic behavior appears, followed by the fully
chaotic regime from $\gamma$ approaching $0.2$. The two systems
become fully synchronized from $\gamma \approx 0.25$ -- in the
bifurcation diagram the difference $x-y$ stays on the zero value,
i.e., the trajectories of the systems $\{X\}$ and $\{Y\}$ are
identical. Then we observe a quasi-symmetry about
 $\gamma = 0.5$, i.e., the synchronized behavior ends for $\gamma
 > 0.75$ and we observe the chaotic, periodic and again chaotic
 behavior of the unsynchronized systems. This development is
 reflected in the mutual information rate (\ref{mirg}), depicted
 in the bottom of Fig.~\ref{log2}. With $\gamma$ increasing from
 zero also the MIR gradually increases, however, it falls down to
 zero for the interval of periodic dynamics. Thus the MIR is not
 simply a measure of dependence of dynamics, it rather quantifies
 an information transfer between systems or processes. In the case
 of periodic systems with the zero entropy rate (KSE), also the
 MIR is zero. In the subsequent chaotic regimes the MIR quickly
 increases with $\gamma$ approaching the synchronization
 threshold. During the fully synchronized regime the MIR stays on
 its maximum value. It is interesting to compare Fig.~\ref{log2}
 with Fig.~4 in Ref.~\cite{blancMIR} where the authors present
 results of their four MIR estimators, stating that the Lempel-Ziv estimator \cite{lempelziv}
and the causal state machine based estimator
\cite{crutch-youngPRL,shalcrutchcompmech,haslinger2010shalizi}
gave the most faithful results. The latter are qualitatively
equivalent to the results obtained using the formula (\ref{mirg})
for the MIR of Gaussian processes, estimated using the FFT (the
bottom graph of Fig.~\ref{log2}). The qualitative equivalence
means that although the values of the MIR estimates are different,
the shapes of the MIR dependence on the coupling parameter
$\gamma$ are very similar.

\begin{figure}
 \includegraphics[width=1.0\columnwidth]{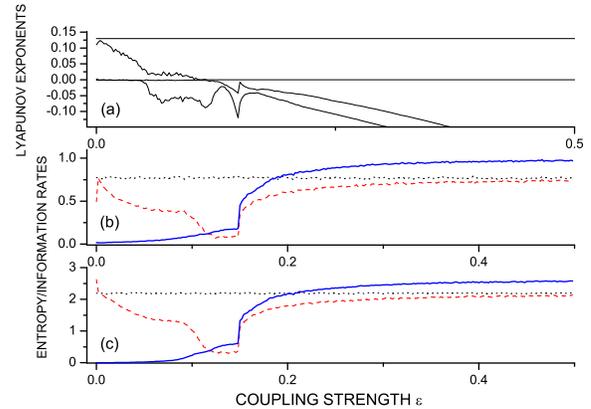}
 \caption{(Color online) (a) Two largest Lyapunov exponents of the drive $\{X\}$
 (the constant lines) and the response $\{Y\}$
 (the decreasing lines), (b) the entropy rates (\ref{hg}) for the drive $\{X\}$
 (the dotted black line) and the response $\{Y\}$ (the dashed red line)
 and the mutual information rate (\ref{mirg}) between $\{X\}$ and
 $\{Y\}$ (the solid blue line) computed using the FFT; (c) the same as
 in (b), but computed using the CCWT;
for the unidirectionally coupled R\"{o}ssler systems
(\ref{aros1}),(\ref{aros2}), as functions of the coupling
strength~$\epsilon$. The Lyapunov exponents are measured in nats
per a time unit; the entropy and information rates are measured in
units of nats per sample. }\label{aros}
 \end{figure}

As an example of a continuous-time system we will consider the
unidirectionally coupled R\"{o}ssler systems, studied also by
Palu\v{s} and Vejmelka \cite{testpap1},
 given by the
equations
\begin{eqnarray}\label{aros1}
\mathaccent95 x_{1} &= &- \omega_{1} x_{2} - x_{3}       \nonumber \\
\mathaccent95 x_{2} &= &\omega_{1} x_{1} + a_1 \; x_{2}    \\
\mathaccent95 x_{3} &= &b_1 + x_{3} (x_{1} - c_1) \nonumber
\end{eqnarray}
for the autonomous system $\{X\}$, and
\begin{eqnarray}\label{aros2}
\mathaccent95 y_{1} &= &- \omega_{2} y_{2} - y_{3} + \epsilon
(x_{1} - y_{1}) \nonumber \\
\mathaccent95 y_{2} &= &\omega_{2} y_{1} + a_2 \; y_{2} \\
\mathaccent95 y_{3} &= &b_2 + y_{3} (y_{1} - c_2) \nonumber
\end{eqnarray}
\noindent for the response system $\{Y\}$. We will use the
parameters $a_1 = a_2 = 0.15$, $b_1 = b_2 = 0.2$, $c_1 = c_2 =
10.0$, and frequencies
 $\omega_{1} = 1.015$ and $\omega_{2} =
0.985$, i.e., the two systems are similar, but not identical.

Figure~\ref{aros}a presents four Lyapunov exponents (LE) of the
coupled systems (the two negative LE's are not shown) as functions
of the coupling strength $\epsilon$. One positive and one zero LE
of the driving  system $\{X\}$ are constant, while the LE's of the
driven  system $\{Y\}$ which are positive and zero without a
coupling or with a weak coupling decrease with increasing
$\epsilon$. The two systems can enter a synchronized regime  when
the originally positive LE of the response system becomes
negative. After a transient negativity and a return to zero, the
originally positive LE of the driven  system $\{Y\}$ becomes
decreasing and negative for $\epsilon > 0.15$ (Fig.~\ref{aros}a).
The mutual information rate (\ref{mirg}) between $\{X\}$ and
 $\{Y\}$  computed using the FFT (the solid blue line in Fig.~\ref{aros}b)
 gradually increases with the increasing coupling strength $ 0 < \epsilon < 0.15$,
 however, shows a steep increase at or after the synchronization
 threshold at $\epsilon \approx 0.15$. Then it again increases
 slowly to its asymptotic value in the synchronized state.
At the coupling strength $\epsilon \approx 0.15$ also the entropy
rate (\ref{hg}) of the driven  system $\{Y\}$ (the dashed red line
in Fig.~\ref{aros}b) steeply increases and then continues in a
gradual increase and asymptotically approaches the entropy rate
(\ref{hg}) of the autonomous system $\{X\}$ (the dotted black line
in Fig.~\ref{aros}b). Due to this behavior Palu\v{s} et
al.~\cite{cir1pre} described the route to synchronization as an
adjustment of information rates. It is important that even the
mutual information rate (\ref{mirg}) of Gaussian processes,
computed using the FFT of time series generated by the studied
systems, reflects both the gradual increase of coupling as well as
the sudden transient into synchronization.

\section{MIR and networks of dynamical systems}\label{mirn}

In Sec.~\ref{secmir} we have introduced the mutual information
rate as a quantity measuring the dependence between dynamics of
two systems or processes. Unlike the static measures such as the
correlation coefficient or the mutual information of random
variables, the estimates of the MIR do not suffer by a bias
dependent on the character of dynamics underlying analyzed time
series. Blanc et al.~\cite{blancMIR} also show that the MIR is
independent of time lag between time evolutions of studied
systems. Together with Blanc et al.~\cite{blancMIR} and Baptista
et al.~\cite{baptistaMIR} we propose the MIR as an association
measure suitable for inferring interaction networks from
multivariate time series generated by coupled dynamical systems.
Specifically in this paper we propose to use the formula
 (\ref{mirg}) for the
 mutual information
rate of Gaussian processes. Although  Gaussian processes are
inherently linear, in Sec.~\ref{gady} we have demonstrated that
the MIR (\ref{mirg}) computed using the FFT of time series
generated by the studied nonlinear dynamical system was able  to
distinguish not only synchronized from unsynchronized states, but
also different levels of dependence between dynamics of the
studied systems due to different strengths of their coupling.
These observations, however, cannot assure a general applicability
of the MIR (\ref{mirg}) for natural nonlinear systems. Before
constructing networks from experimental multivariate time series
it is necessary to test for a presence of nonlinearity in studied
time series and assess its actual effect on the inference and
quantification of dependence relations present in the data. It is
not surprising that such studies have been done in the same areas
where the research based on the complex networks paradigm is very
active.

Functional brain networks are frequently constructed using time
series from sequences of functional magnetic resonance imaging
(fMRI) \cite{Bullmore2009,achard06}. Hlinka et
al.~\cite{hlinkalinfmri} demonstrate that the linear correlation
coefficient is a sufficient measure of functional connectivity in
resting-state fMRI data. Potential new information brought by
nonlinear measures such as the mutual information is relatively
minor and negligible in comparison with natural intra- and
inter-subject variability. Hartman et al.~\cite{hartman} confirm
this finding in specific computations of graph-theoretical
measures from fMRI brain networks. Also spatio-temporal dependence
structures in electrophysiological data such as the
electroencephalogram (EEG) are characterized within the complex
networks paradigm \cite{Bullmore2009,stam07}. Nonlinear character
of the EEG in epilepsy is known~\cite{Casdagli1996381}, some level
of nonlinearity can be detected also in normal human EEG
recordings~\cite{rdbc}. Distinction of different physiological
and/or pathological brain states observed using nonlinear measures
can successfully be reproduced by a proper application od standard
tools derived from the theory of linear stochastic processes
\cite{theirapp}. While the latter findings characterized
single-channel EEG signals, the character of dependence between
EEG signals from different parts of the scalp are relevant for the
construction of the EEG brain networks. Nonlinear measures have
been applied in order to distinguish different consciousness
states using so-called multichannel attractor embedding
\cite{matousek1995global}. Changes in dependence structures in
multichannel EEG data which have been described by a nonlinear
measure such as the correlation dimension from the multichannel
embedding \cite{matousek1995global}, however, can be equivalently
captured by a linear measure extracted from a correlation matrix
\cite{spat0}.

Using an equivalent approach, climate networks
\cite{tsonis2004architecture,tsonis2006,yamaninoprl,yamasaki09,donges2009,donges2009backbone,gangclidym}
are constructed using multivariate time series of long-term
records of meteorological variables such as the air temperature or
pressure.
Already in the 1980's a number of researches attempted to infer
nonlinear dynamical mechanisms from meteorological data and
claimed detections of a weather or climate attractor of a low
dimension \cite{nicolis84, fraedrich86, tsonis88}. Other authors
pointed to a limited reliability of chaos-identification
algorithms and considered the observed low-dimensional
weather/climate attractors as spurious
\cite{grassberger86,lorenz91}. Palu\v{s} \& Novotn\'a \cite{meteo}
even found the air temperature data well-explained by a linear
stochastic process, when the dependence between a temperature time
series $\{x(t)\}$ and its lagged twin $\{x(t + \tau)\}$ was
considered. Hlinka et al.~\cite{hlinkaclidy} extended the later
result to the dependence between the monthly time series of the
gridded whole-Earth air temperature reanalysis data. These results
do not mean that the dynamics underlying records of meteorological
data is linear. For instance,
a search for repetitive patterns on specific temporal scales in
the air temperature and other meteorological data has led to an
identification of oscillatory phenomena possibly possessing a
nonlinear origin and exhibiting phase synchronization between
oscillatory modes extracted either from different types of
climate-related data or data recorded at different locations on
the Earth \cite{npgsvd,npgqbo,jastp2,feliks10,pano11}. The studies
of Hlinka et al.~\cite{hlinkaclidy,hlinkaentro13} merely state
that for inferring general dependence and causal relations, the
approaches derived for Gaussian processes perform very well and
nonlinear approaches do not bring substantial new information.

These arguments and the fact that the mutual information rate
estimator, computed using the FFT and the formula (\ref{mirg}) for
the MIR of Gaussian processes is computationally less demanding
that estimators for general nonlinear processes, form the basis
for our recommendation of the MIR (\ref{mirg}) as a measure
suitable for inference of networks from experimental multivariate
time series recorded from complex systems of various origins.
There is still a serious demand for the amount and stationary
character of the analyzed data, since the computation of the
magnitude-squared coherence (\ref{gamma}) is based on dividing the
time series into a number of segments over which the complex
cross-spectrum (the numerator in the Eq. (\ref{gamma}) right-hand
side) is averaged. In many cases time series from natural complex
systems are relatively short and nonstationary. Nonstationarity in
the sense of changing relationships between time series with time
leads to changes in the strength and even the existence of links
in interaction networks during some time intervals. The complex
network paradigm copes with this phenomenon using the concept of
temporal networks \cite{temporalnets} or evolving networks. The
latter approach assumes approximate step-wise stationarity of the
analyzed time series and a standard ``static'' network is inferred
in a relatively short time window which is ``sliding'' over the
whole time interval spanned by the available experimental time
series. The time evolution of graph-theoretical characteristics is
then studied with respect to a time evolution and/or an occurrence
of marked events in the studied complex system. This approach has
been successfully applied in the EEG brain networks
\cite{lehnertz-evo1,lehnertz-evo2,lehnertz-evo3}, as well as in
the climate networks \cite{radebachPRE}. An alternative approach,
applied in the field of climate networks, is ``picking-up'' a
number of unequal-length subsets of the whole time series, tight
to an occurrence of some phenomenon (e.g. El Ni\~{n}o) and
performing the summation in the formula (\ref{cxy}) for the
correlation coefficient only using the selected subsets of the
data \cite{tsonis2008prl}. Neither the latter approach, nor the
evolving network strategy can be applied when using the standard
FFT-based evaluation of the MIR (\ref{mirg}).

In order to cope with nonstationarity we propose to use a wavelet
transform instead of the Fourier transform. In particular, the
complex continuous wavelet transform (CCWT) is applied in order to
convert a time series $x(t)$ into a set of complex wavelet
coefficients $W(t,f)$:

\begin{equation}
W(t,f)=\int_{-\infty}^{\infty}\psi(t')x(t-t')dt'\label{eq:wavelet_transform}\end{equation}
using the complex Morlet wavelet \cite{ccwt}:

\begin{equation}
\psi(t)=\frac{1}{\sqrt{2\pi\sigma_{t}^{2}}}\exp(-\frac{t^{2}}{2\sigma_{t}^{2}})\exp(2\pi
if_{0}), \label{eq:Morlet}\end{equation}
where  $\sigma_{t}$ is the bandwidth parameter, and $f_{0}$ is the
central frequency of the wavelet. $\sigma_{t}$ determines the rate
of the decay of the Gauss function, its reciprocal value
$\sigma_{f}=1/\pi\sigma_{t}$ determines the spectral bandwidth. In
order to keep the wavelet representation close to the original MIR
(\ref{mirg}) evaluation based on the FFT, we use a set of
equidistantly spaced central wavelet frequencies in the relevant
frequency range given by the time series length and its sampling
frequency, instead of the power-law pyramidal scheme, usually used
in the wavelet context. Then the product of the complex wavelet
coefficients $W_X(t,f)W_Y ^*(t,f)$,
as well as the norms $|W_X(t,f)|$, $|W_Y(t,f)|$ are
 averaged over time $t$. Finally, the wavelet magnitude-squared
 coherence

\begin{equation}\label{gammaw}
|\Gamma_{X,Y}^W (f)|^2 = \frac{|W_{X,Y}
(f)|^2}{|W_{X}(f)||W_{Y}(f)|}
\end{equation}
is used in the summation over the set of the central wavelet
frequencies according to Eq.~(\ref{mirg}). Here $W_{X,Y} (f)$,
$|W_{X}(f)|$, and $|W_{Y}(f)|$ stand for the time averages of
$W_X(t,f)W_Y ^*(t,f)$, $|W_X(t,f)|$, and $|W_Y(t,f)|$,
respectively.

Let us return to the unidirectionally coupled R\"{o}ssler systems
(\ref{aros1}), (\ref{aros2}). Using the wavelet representation we
can recompute both the mutual information rate (\ref{mirg}) and
the entropy rate (\ref{hg}) as functions of the coupling strength
$\epsilon$ (Fig.~\ref{aros}c) . While the CCWT-based estimators
give different values than the FFT-based estimators, they agree in
the qualitative sense that the curves of the $\epsilon$-dependence
of the MIR in Figs.~\ref{aros}b and \ref{aros}c (solid/blue
curves) are the same. The two estimators also give a good
agreement in the $\epsilon$-dependence of the entropy rates
(dashed/red curves in Figs.~\ref{aros}b and \ref{aros}c), there is
just a small difference in the entropy rates of the driven system
for very small values of $\epsilon$.

It is important that the wavelet-based estimator of the MIR
(\ref{mirg}) gives a relative distinction of coupling regimes with
different coupling strengths. This is a property which we expect
from an association measure suitable for the inference of
interaction networks from multivariate time series. It will assign
proper weights to network edges, an edge of two more strongly
coupled nodes (dynamical systems) will obtain a greater weight
than edges connecting nodes with a weaker coupling. For the
construction of the binary networks, a greater value of MIR for
strongly coupled nodes assure an existence of an edge by exceeding
a chosen threshold or a critical value given by a statistical
test. For establishing statistically significant links we propose
to use the surrogate data strategy as described in
Refs.~\cite{testpap1,contempap}. The temporal averaging of the
product of the complex wavelet coefficients $W_X(t,f)W_Y ^*(t,f)$
might evoke a temptation to randomize the phases $\varphi$ of the
complex wavelet coefficients $W(t,f) \equiv A(t,f) \exp \big(i
\varphi(t,f)\big)$. Generating the FFT-based surrogate data
\cite{Theiler1992}, the set of the original phases of the Fourier
coefficient is substituted by a set of independent, identically
distributed (IID) phases randomly sampled from a uniform
distribution on the interval $(0, 2\pi)$. However, the phase
differences of the wavelet coefficients of two signals are not
IID, even if the underlying processes are independent. Using
random IID phases in the summation of $W_X(t,f)W_Y ^*(t,f)$ would
underestimate the critical values in the test for independence and
false edges would be inferred. The character of the (long-range)
dependence of the phase differences in $W_X(t,f)W_Y ^*(t,f)$ of
independent processes depends on the central wavelet frequency.
Therefore, it is more convenient to generate surrogate data and
estimate the MIR from them as in the usual surrogate data test
strategy \cite{testpap1,contempap}.

Unlike in the FFT-based MIR estimation, we apply the CCWT on the
whole time interval of available data. Then we either average the
product of the complex wavelet coefficients $W_X(t,f)W_Y ^*(t,f)$,
as well as the norms $|W_X(t,f)|$, $|W_Y(t,f)|$,
 over the whole time interval or
apply the sliding-window strategy of the evolving networks
\cite{radebachPRE} or the strategy of Tsonis and Swanson
\cite{tsonis2008prl} of the averaging over time intervals selected
according to an occurrence of a specific phenomenon. Using the
strategies that cope with nonstationarity, however, one should
consider a smoothing effect of the wavelet coefficients for low
frequencies (large time scales). Since time series from natural
complex systems frequently reflect processes with a $1/f$
spectrum, the wavelet coefficients for low frequencies have much
greater weights than the coefficients for high frequencies and
effects of short-living phenomena can be masked in the resulted
MIR estimates. Therefore we recommend to limit the final summation
in the MIR formula (\ref{mirg}) to higher frequencies or shorter
time scales in which the effect of short-living phenomena is not
attenuated. The latter idea can be generalized and even for
stationary time series one can restrict the MIR evaluation to a
specific range of time scales, i.e. to a specific spectral band.
Then a scale-specific or frequency-specific connectivity is
evaluated and {\it scale-specific} or {\it frequency-specific
interaction networks} can be studied.

Until now we have considered the MIR $i(X_{i}; Y_{i})$ of two
stochastic processes $\{X_{i}\}$, $\{Y_{i}\}$. Constructing a
network of $n$ nodes, i.e., $n$ dynamical systems, we will
consider $n$ time series as realizations of $n$ stochastic
processes $\{X^k _{i}\}$, $k=1, \dots, n$. (For simplicity we
consider $n$ univariate time series/stochastic processes, an
equivalent of a multivariate stochastic process with $n$
components. The considerations here can be generalized to $n$
multivariate stochastic processes with various numbers $n_i$ of
components.) Then we can evaluate the standard bivariate MIR
$i(X^k _{i}; X^l _{i})$ for each pair of components. In order to
distinguish direct from indirect interactions we can also
consider conditional (partial)
 MIR $i(X^k _{i}; X^l_{i} | X^j_{i}; j=1, \dots, n, j \neq k, j \neq
 l)$ which quantifies the ``net'' dependence between the two
 processes without an influence of the remaining $n-2$ processes.

 For the evaluation of the conditional MIR, in the framework of
 Gaussian processes, we will follow the work of Schelter et
 al.~\cite{schelterpartial} who extended the notion of partial
 correlations to the partial mean phase coherence.

For each pair of processes $\{X^k _{i}\}$, $\{X^l _{i}\}$ and each
central wavelet frequency $f \in \{f_1, f_2, \dots, f_{N_f} \}$ we
evaluate the time-averaged complex wavelet coherence

\begin{equation}\label{gammawcompl}
\Gamma_{k,l}^W (f) = \frac{W_{k,l}
(f)}{\sqrt{|W_{k}(f)||W_{l}(f)|}}.
\end{equation}
Thus for each $f$ we obtain a complex $n\times n$ matrix $\Gamma^W
(f)$. This complex matrix is inverted, $\Omega(f) = (\Gamma^W
(f))^{-1}$. Using the entries of the inverted complex matrix
$\Omega(f)$ we evaluate the conditional wavelet coherence as

\begin{equation}\label{conwavcoh}
\Upsilon_{k,l} (f) = \frac{\Omega_{k,l}
(f)}{\sqrt{|\Omega_{k,k}(f)||\Omega_{l,l}(f)|}}.
\end{equation}

Finally, the magnitude-squared conditional wavelet coherence is
used in the summation according to Eq.~(\ref{mirg}) in order to
obtain the conditional MIR

$$i_G(X^k _{i}; X^l_{i} | X^j_{i}; j=1, \dots, n, j \neq k, j \neq
 l) = $$

\begin{equation}\label{condmirg}
 - \frac{1}{2N_f} \sum_{f=f_1} ^{f_{N_f}} \log(1 - |\Upsilon_{{k,l}}
(f)|^2).
\end{equation}


\section{Climate networks}\label{clinet}

Understanding the complex dynamics of the Earth atmosphere and
climate is a great scientific challenge with a potentially high
societal impact. In their seminal paper, Tsonis and Roebber
\cite{tsonis2004architecture} have proposed to study the climate
system as a complex network. Since then the field of climate
networks is rapidly developing and expanding in the scope of
methodology as well as applications.
The spatio-temporal dynamics of the atmosphere is captured by
multivariate time series of long-term recordings of meteorological
variables. Typically, such instrumental data are preprocessed and
interpolated in order to assign a time series of a variable to
each node of a regular angular grid covering the Earth surface, as
well as slices of the atmosphere at various altitude or air
pressure levels. Such gridded time series of meteorological
variables, available due to, e.g.,  the NCEP/NCAR reanalysis
project \cite{ncep} are usually, although not exclusively, used
for the construction of climate networks.
 Monthly \cite{tsonis2008prl,donges2009} or daily
\cite{yamaninoprl,yamaninoepl} surface air temperature data are
frequently used, however, equipotential heights
\cite{tsonisteleconnection,dongeshladiny}, sea surface
temperature, humidity,
 precipitation and related data \cite{gangclidym,malikmarwan}  and other
meteorological data are also analysed. Individual grid-points,
characterized by time series of a chosen meteorological variable,
are considered as nodes (vertices) of a climate network, while
links (edges) are inferred from some, mostly statistical
association between the time series related to the two nodes at
the edge's end-points. The most common association measure is the
Pearson's correlation coefficient
\cite{tsonis2004architecture,tsonis2008prl}, however, also the
Spearman's rank correlation coefficient is used \cite{tropevolv},
and more general, nonlinear measures are tested, e.g., the
bivariate mutual information \cite{donges2009,donges2009backbone},
and the mutual information of ordinal time series
\cite{latinochaos,latinoepj} or measures from the phase
synchronization analysis \cite{yamaphasyn} and the event
synchronization analysis \cite{malikmarwan}.

In the following study we use monthly mean values of the
near-Surface Air Temperature (SAT) from the NCEP/NCAR reanalysis
\citep{ncep}. We include the data up to the latitudes
87.5$^{\circ}$ in the grid of 2.5$^{\circ}$x2.5$^{\circ}$ which
leads to 10,224 grid points or network nodes.
 The temporal interval of 624 months starting in January 1958
 and ending in December 2009 is used for the inference of the network
 using the correlation coefficient and the CCWT-based MIR
 estimator. For the FFT-based estimator the temporal interval is
 extended backward by 16 months (starting in September 1956) in
 order to have 5 segments of 128 monthly samples.

\begin{figure*}
\begin{center}
 \includegraphics[width=0.9\linewidth,height=0.5\linewidth]{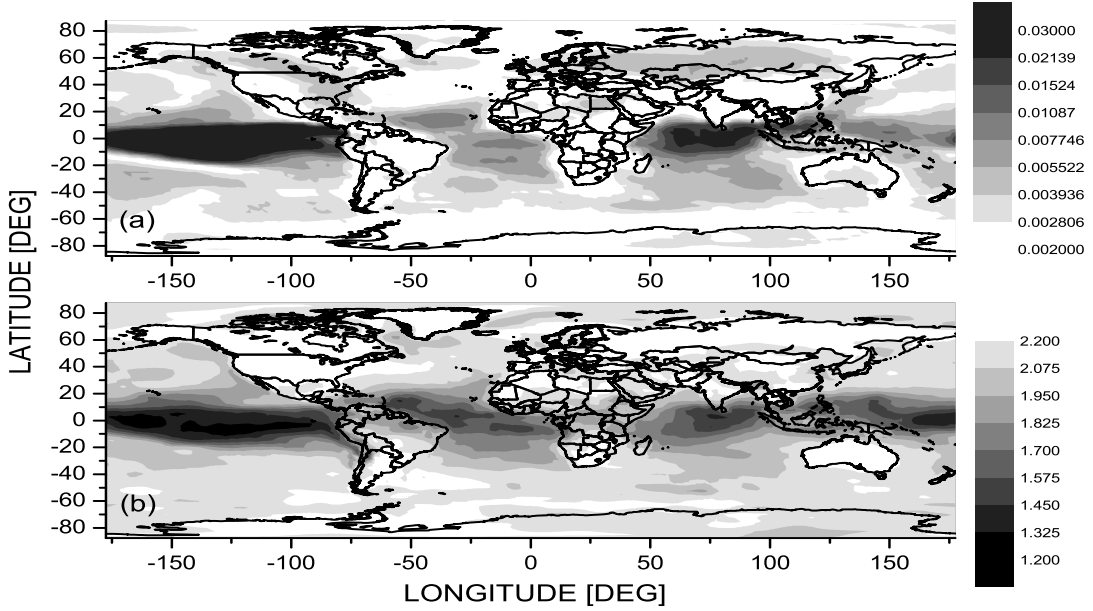}
\caption{(a) Area weighted connectivity for the SATA climate
network with the density $\varrho = 0.005$ obtained by the uniform
thresholding of the absolute correlations. (b) Complexity of each
node SATA time series measured by the Gaussian process entropy
rate $h_G$. Note the reversed grey scales, the black color
corresponds to the largest AWC in (a), while in (b) it corresponds
to the lowest entropy rates. } \label{acawcgser}
\end{center}
\end{figure*}

 In order to avoid trivial correlations due to seasonal
 temperature variability, the annual cycle has been removed from
 each SAT time series. The SAT anomalies (SATA thereafter) have
 been computed by subtracting the averages for each month from
 related samples, e.g., the average January temperature was
 subtracted from all January samples, etc.

As the first step of the network analysis we compute the
correlation coefficients $c_{i,j}$ for each pair of nodes $i,j =
1, \dots, N_N=10224$. We use the matrix of the absolute
correlations $C_{i,j} = |c_{i,j}|$ in order to obtain the
adjacency matrix $A_{i,j}$ of the binary network, defined as:
$A_{i,j} =1 $ iff $C_{i,j}>c_T$, otherwise $A_{i,j}=0$.
$A_{i,i}=0$ by definition. The total number of existing edges
divided by the number of all possible edges in known as the
network density (or edge density) $\varrho$. Following Donges et
al. \cite{donges2009,donges2009backbone} we choose the threshold
$c_T$ such that the resulting network density is $\varrho =
0.005$.

The basic characterization of connectivity of a node $i$ is its
degree, or degree centrality $k_i$
\begin{equation}\label{degree}
k_i = \sum_{j=1}^{N_N} A_{i,j},
\end{equation}
giving the number of nodes to which the node $i$ is connected.
Since the reanalysis data are defined on a grid which is regular
in the angular coordinates, the geographic distances of the grid
points depend on the latitude $\lambda_i$. In order to correct for
this dependence, for the climate networks defined on the regular
angular grid  the area weighted connectivity \cite{tsonis2006} is
defined as
\begin{equation}\label{awc}
AWC_i =\frac{ \sum_{j=1}^{N_N} A_{i,j} \cos(\lambda_j)}{
\sum_{j=1}^{N_N} \cos(\lambda_j)}.
\end{equation}
The AWC   can be interpreted as the fraction of the Earth's
surface area a vertex is connected to.

The AWC computed for each node of the SATA network based on the
absolute correlations  with $\varrho = 0.005$ (AC-network in the
following) is mapped in Fig.~\ref{acawcgser}a. According to this
analysis the most connected nodes (``hubs'') of the climate
network lie in the tropical areas of the Pacific and Indian
Oceans. The hub in the tropical Pacific include so-called El
Ni\~{n}o areas.
The El Ni\~{n}o/Southern Oscillation (ENSO) is a dominant mode of
the global atmospheric circulation variability which
quasiperiodically causes shift in winds and ocean currents
centered in the Tropical Pacific region and is linked to anomalous
weather/climate patterns worldwide \cite{sarachik2010nino}.
The global influence of the El Ni\~{n}o phenomenon is used to
explain the observation that the El Ni\~{n}o area constitutes the
principal hub of the climate network.

\begin{figure}
 \includegraphics[width=1.0\columnwidth]{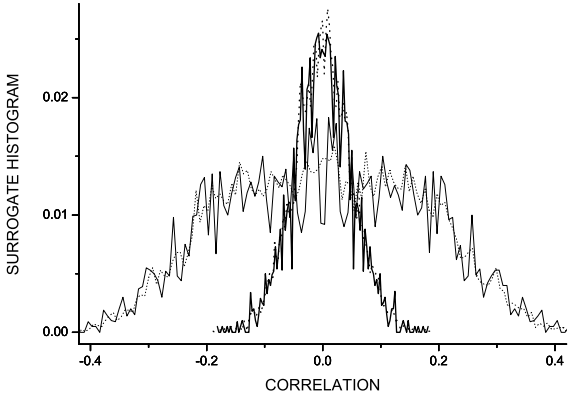}
 \caption{Histograms
of cross-correlations of  independent FT (solid lines) and
amplitude-adjusted FT (dotted lines) surrogate data for the SATA
of a pair of nodes from the low entropy rate area (the thin lines,
the nodes with the latitude 0$^{\circ}$, the longitude
90$^{\circ}$W and 10$^{\circ}$S, 120$^{\circ}$W) and a pair from
the high entropy rate area (the thick lines, the nodes
60$^{\circ}$N, 25$^{\circ}$E and 60$^{\circ}$N, 75$^{\circ}$E).
 }\label{surrhist}

 \end{figure}

Let us characterize the dynamics of the SATA time series using the
entropy rate (\ref{hg}). The FFT-based estimator assign an entropy
rate value to each grid-point, i.e. to each node of the network,
so they can be mapped in the same way as the AWC. The entropy rate
map is presented in Fig.~\ref{acawcgser}b. The correspondence
between the lowest entropy rates and the highest AWC of the
AC-network is indisputable. In order to assess a bias in the
correlation estimator we need time series which are independent,
but have the same dynamics (the same entropy rate) as the original
SATA time series. Such time series can be generated using the
FFT-based surrogate data algorithm \cite{Theiler1992,contempap}.
The fast Fourier transform is applied to a time series, the
magnitudes of the complex Fourier coefficients are preserved, but
their phases are randomized. Using different sets of random phases
the inverse FFT generates a number of independent realizations of
a Gaussian process with the same spectrum (and thus with the same
entropy rate (\ref{hg})) as that of the original time series. A
potential digression from the Gaussian distribution is solved by a
histogram transformation known as the amplitude adjustment. We use
both the FT surrogate data and the amplitude-adjusted FT (AAFT)
surrogate data, however, they give equivalent results. Generating
a large number of realizations of the FT (AAFT) surrogate data for
the SATA time series we can estimate distributions of the
correlations of independent surrogates of the SATA series from
various grid-points. Figure~\ref{surrhist} compares such histogram
for SATA-surrogates for a pair of grid-points from a low entropy
rate area (the El Ni\~{n}o area) and from a high entropy rate area
(an Euro-Asian area on 60$^{\circ}$N). While in the Northern
hemisphere high entropy rate area the correlation bias (the
cross-correlation of realizations of independent processes)
scarcely reaches over $\pm 0.1$, in the tropical Pacific areas the
cross-correlation bias can reach values close to $\pm 0.4$.

\begin{figure*}
\begin{center}
 \includegraphics[width=0.9\linewidth,height=0.5\linewidth]{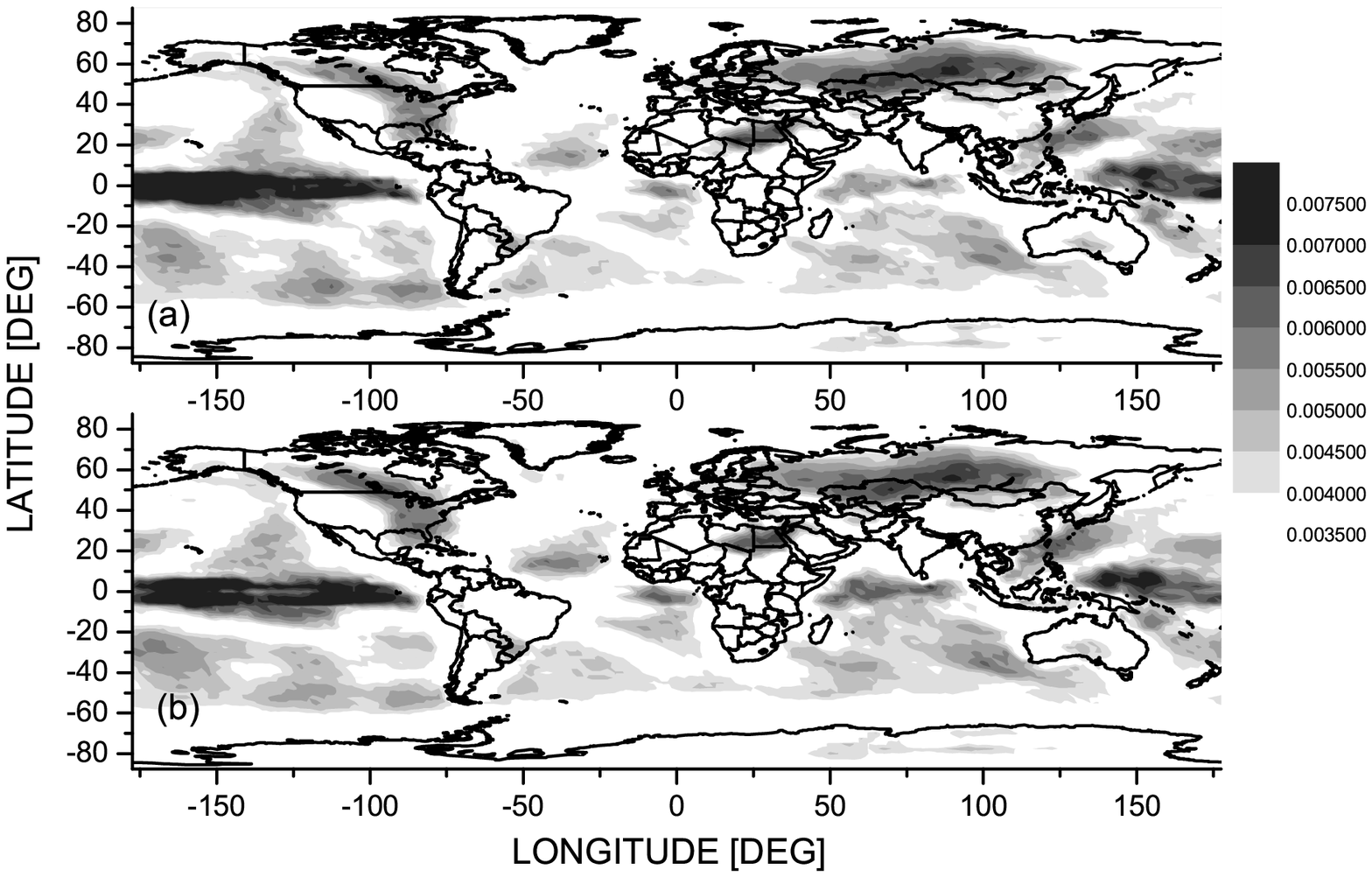}
\caption{(a) Area weighted connectivity for the SATA climate
network with the density $\varrho = 0.005$ obtained by the uniform
thresholding of the mutual information rate (\ref{mirg}) estimated
using the Fourier transform (a) and the continuous complex wavelet
transform (b). Note that the scale is different from that in
Fig.~\ref{acawcgser}a.} \label{awcrtftwt}
\end{center}
\end{figure*}

\begin{figure}
 \includegraphics[width=1.0\columnwidth]{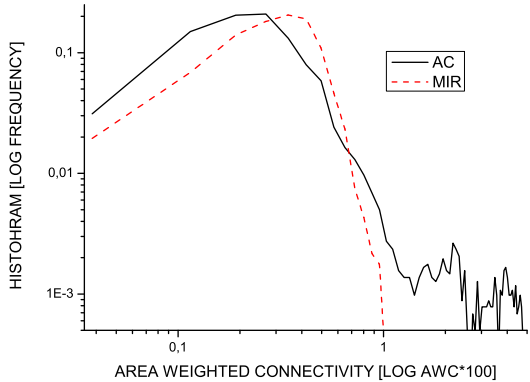}
 \caption{(Color online) Histograms (64 bins)
 of the area weighted connectivity (AWC*100) for the SATA climate network
with the density $\varrho = 0.005$ obtained by the uniform
thresholding of the absolute correlations (solid black line), and
of the mutual information rate (\ref{mirg}) (dashed red line).
 }\label{awchist}

 \end{figure}

\begin{figure}
 \includegraphics[width=1.0\columnwidth,height=0.5\linewidth]{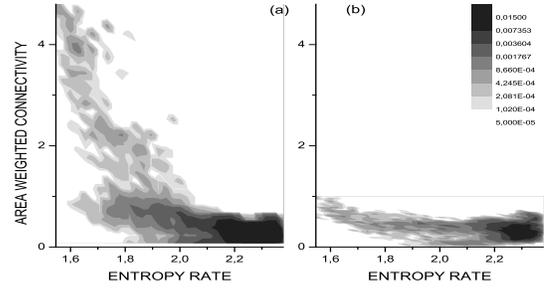}
 \caption{(a) Grey-coded bivariate histograms (32x32 bins)
 reflecting the joint probability distribution of the SATA
 entropy rate (\ref{hg}), cf. Fig.~\ref{acawcgser}b,
 and the area weighted connectivity (AWC*100) for the SATA climate network
with the density $\varrho = 0.005$ obtained by the uniform
thresholding of the absolute correlations, cf.
Fig.~\ref{acawcgser}a. (b) The same as in (a), but considering the
area weighted connectivity for the SATA climate network with the
density $\varrho = 0.005$ obtained by the uniform thresholding of
the mutual information rate (\ref{mirg}), cf.
Fig.~\ref{awcrtftwt}b.
 }\label{erawchst2}

 \end{figure}

As an alternative we construct a climate network using the MIR
(\ref{mirg}) and again we threshold the MIR values in order to
obtain the network density  $\varrho = 0.005$. The area weighted
connectivity for the MIR-networks is mapped in
Fig.~\ref{awcrtftwt}, where we can compare the results for both
the FFT- and CCWT-based estimators. The results are quite similar.
A few small differences can be caused by the fact that the
FFT-based estimator used segments of 128 samples and thus cannot
include the connectivity on large time scales as the CCWT
estimator which utilizes 624 samples in one whole segment. The
differences between the MIR-network (Fig.~\ref{awcrtftwt}) and the
AC-network (Fig.~\ref{acawcgser}a) are much larger and more
important. In the comparison with the AC-network, the very
connected hub in the Indian Ocean almost disappears in the MIR
network. The hub in the El Ni\~{n}o area survives, however, it is
weaker and confined to a smaller area. The connectivity in the
continental areas of the Northern hemisphere increases in the
MIR-network. This comparison, however, cannot give an answer which
network representation is closer to the physical reality.

In the analogy with degree distributions, studied in the complex
network theory, in Fig.~\ref{awchist} we present histograms
estimating the distributions of the area weighted connectivity.
The AWC distribution for the AC-network (the solid black line)
shows a heavy irregular tail of extreme AWC values, while the AWC
distribution for the MIR-network (the dashed red line) shows a
distribution bounded by a fast probability decay for large AWC
values, well captured by a Poisson distribution.
Scholz~\cite{nodesim} obtained such distributions using a
node-similarity network model. Each node has a set of features,
quantified as coordinates in an Euclidean space. Based on a random
data set, two nodes are defined as connected (similar) when their
Euclidean distance is below a certain threshold. Using a small
threshold only very similar (close) nodes are connected. This
represents a sparsely connected network showing typically
scale-free power-law like distributions. Increasing the threshold,
more densely connected networks are modelled with node degree
distributions very similar to that of the MIR-network (the dashed
red line in Fig.~\ref{awchist}).

\begin{figure*}
\begin{center}
 \includegraphics[width=0.9\linewidth,height=0.5\linewidth]{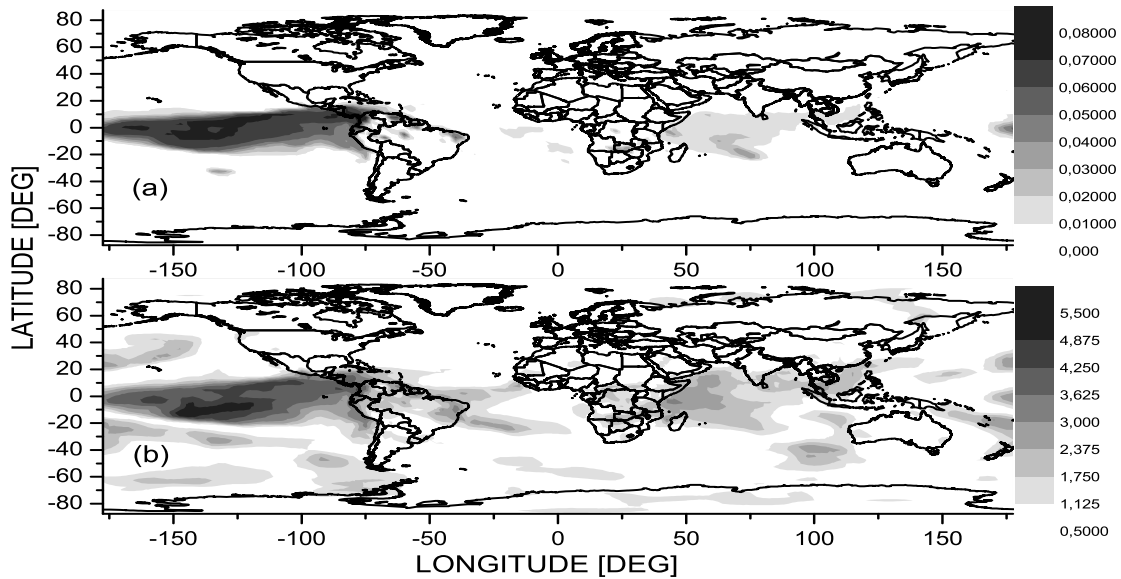}
\caption{(a) Area weighted connectivity for the scale-specific
SATA climate network with the density $\varrho = 0.005$ obtained
by the uniform thresholding of the mutual information rate
(\ref{mirg}) estimated using  the continuous complex wavelet
transform within the scales related to the periods 4--6 years. (b)
Dependence of the SATA time series on the Southern Oscillation
index measured by MIR (\ref{mirg}) estimated using  the CCWT
within the scales related to the periods 4--6 years. }
\label{enso4-6}
\end{center}
\end{figure*}

Bivariate histograms estimating the joint probability distribution
of the SATA entropy rate and the AWC for the studied climate
networks are presented in Fig.~\ref{erawchst2}. In the AC-network
the extremely high AWC values are tight to the nodes with the low
entropy rate of the SATA time series (Fig.~\ref{erawchst2}a).
Together with the histograms in Fig.~\ref{awchist} this picture
supports the conclusion that the very high AWC values of the nodes
characterized by low entropy rates are probably consequences of
the bias in the absolute correlation estimations. The MIR-network
 lacks extreme AWC values, however, some
tendency for the preference of higher AWC values in the nodes with
low entropy rates remains (Fig.~\ref{erawchst2}b). Since the MIR
should not be biased upward by the low entropy rate, this
dependence reflects the physical reality: The nodes in the El
Ni\~{n}o area are the hub of the climate networks. Their increased
connectivity reflects distant influences of the ENSO phenomenon.
On the other hand, the quasiperiodic ENSO behavior in certain
frequency ranges increases the dynamical memory/decreases the
entropy rate of the SATA time series in the El Ni\~{n}o area. We
will demonstrate these phenomena using the scale-specific
connectivity in the next Section.

\section{Scale-specific climate networks}\label{ssclin}

Using the idea of the scale-specific connectivity reflected by the
CCWT-based MIR estimates in which the summation over the wavelet
scales (central wavelet frequencies) is restricted to a chosen
scale range (Sec.~\ref{mirn}) we will study scale-specific SATA
climate networks. Starting with the MIR estimate restricted to the
wavelet time scales corresponding to the periods 4--6 years, in
Fig.~\ref{enso4-6}a we map the AWC for the scale-specific SATA
climate network for the time scales 4--6 years (SSCN(4--6yr)
thereafter). As in the previous cases we consider the binary
network with $\varrho = 0.005$. The hub of this network, i.e., the
highest scale-specific connectivity in the time scales 4--6 years
is located in the tropical Pacific area. It is not surprising
since the oscillatory modes in the range of quasi-biennial
oscillations (QBO, periods 2--3 years) and quasi-quadrennial
oscillations (QQO, the periods fluctuating between 3 and 7 years)
have been detected in the quasi-periodic ENSO dynamics
\cite{ensoqqqbo,kondrashov2005hierarchy}. The scale-specific
MIR-network for the QBO scale 2--3 years (not presented) has
practically the same AWC geographical distribution as the
MIR-network for the used QQO range 4--6 years
(Fig.~\ref{enso4-6}a). It is interesting to note that the SATA
oscillatory mode with the periods 2--3 years is not simply a
higher harmonic \cite{harmonics} of the mode with the periods 4--6
years, since the test for the 1:2 phase coherence between these
modes did not reject the null hypothesis of phase-independence of
these oscillatory modes.

\begin{figure*}
\begin{center}
 \includegraphics[width=0.9\linewidth,height=0.5\linewidth]{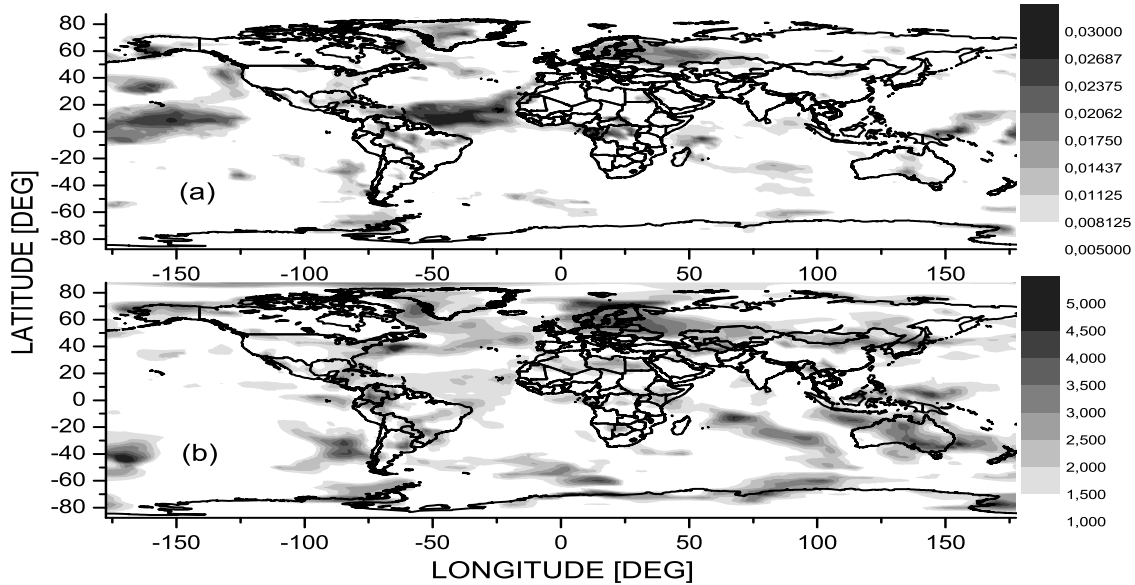}
\caption{(a) Area weighted connectivity for the scale-specific
SATA climate network with the density $\varrho = 0.005$ obtained
by the uniform thresholding of the MIR (\ref{mirg}) estimated
using the CCWT within the scales related to the periods 7--8
years. (b) Dependence of the SATA time series on the North
Atlantic Oscillation index measured by MIR (\ref{mirg}) estimated
using the CCWT within the scales related to the periods 7--8
years. } \label{nao7-8}
\end{center}
\end{figure*}

The ENSO is characterized by several indices derived from the sea
surface temperature and the Southern Oscillation index (SOI, see
http://www.cru.uea.ac.uk/cru/data/soi/ for the data and their
description) which  is defined as the normalized air pressure
difference between Tahiti and Darwin. The MIR quantifying the
dependence within the time scales 4--6 years between the SOI and
the SATA time series in each grid-point is illustrated in
Fig.~\ref{enso4-6}b. The hubs of the SSCN(4--6yr) in the tropical
Pacific and Indian Oceans (Fig.~\ref{enso4-6}a) are parts of the
areas connected to the ENSO within this time scale
(Fig.~\ref{enso4-6}b). However, the areas connected to the ENSO
are quite more extended in the Pacific Ocean, tropical Atlantic
Ocean and in the Indian and Southern Ocean. Also large continental
areas in the Central and Southern America, areas in Africa and
some areas in Asia and Northern America have the SATA variability
in the time scale 4--6 years connected to the ENSO. This extended
ENSO scale-specific connectivity is apparently reflected also in
the broad-band connectivity
and confirms the role of the hub of the global climate networks
for the ENSO tropical Pacific area, as we have observed in the
previous Section. The quasi-periodic dynamics plays an important
role in the ENSO area temperature variability, e.g. the QQO mode
explains almost 40\% of the variability of the sea surface
temperature anomalies \cite{ensoqqqbo}. This fact explains the low
entropy rate of the SATA time series in this area and the
dependence between the AWC and the entropy rate in
Fig.~\ref{erawchst2}b.

\begin{figure*}
\begin{center}
 \includegraphics[width=0.9\linewidth,height=0.5\linewidth]{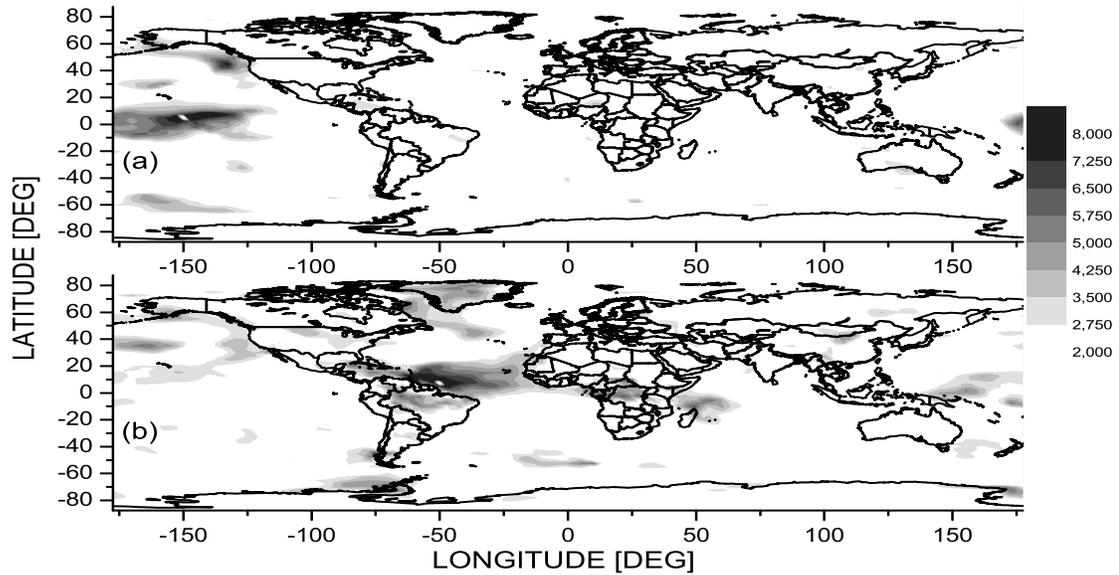}
\caption{(a) Dependence of the SATA time series from each node
with the SATA time series in the node with the longitude
150$^{\circ}$W and the latitude 5$^{\circ}$N, measured by MIR
(\ref{mirg}) estimated using  the CCWT within the scales related
to periods 7--8 years. (b) The same as in (a), but for the node
50$^{\circ}$W, 7.5$^{\circ}$N. The reference node can be seen as a
white pixel in the black background. } \label{pointNAO78}
\end{center}
\end{figure*}

\begin{figure*}
\begin{center}
 \includegraphics[width=0.9\linewidth,height=0.5\linewidth]{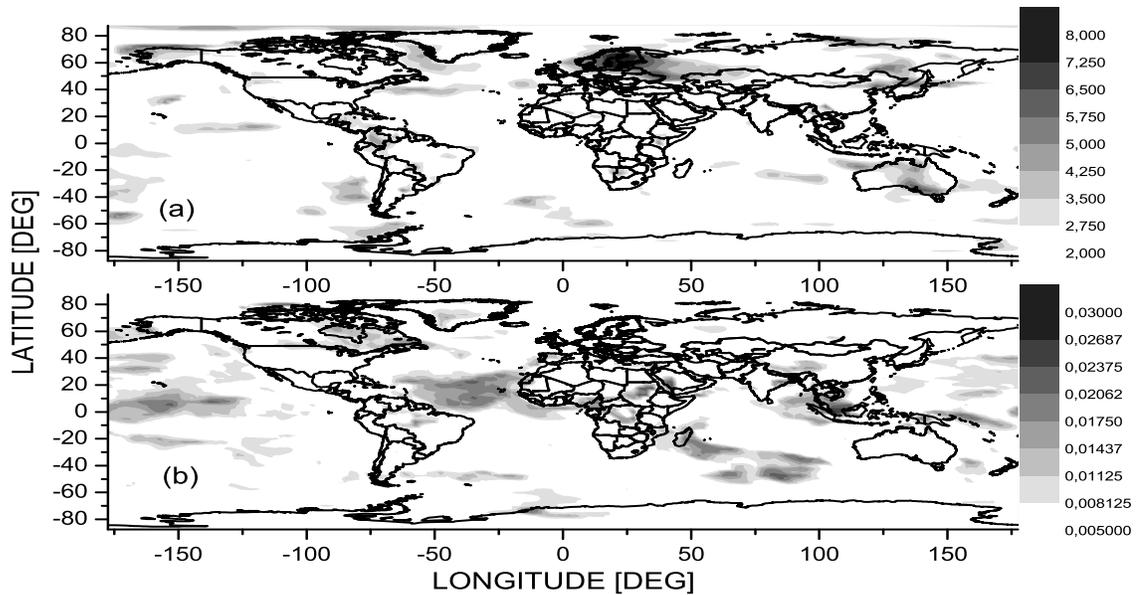}
\caption{(a) Dependence of the SATA time series from each node
with the SATA time series in the node 22.5$^{\circ}$E,
60$^{\circ}$N, measured by MIR (\ref{mirg}) estimated using  the
CCWT within the scales related to periods 7--8 years. (b) Area
weighted connectivity for the conditional scale-specific SATA
climate network with the density $\varrho = 0.005$ obtained by the
uniform thresholding of the MIR (\ref{mirg}) estimated using  the
CCWT within the scales related to the periods 7--8 years. The MIR
between each two nodes is taken conditionally on the NAO index. }
\label{pointNAOCNAO78}
\end{center}
\end{figure*}

Oscillatory phenomena with the  period around 7--8 years have been
observed in the air temperature and other meteorological data by
many authors (see Ref.~\cite{jastp2,pano11} and references
therein).
Therefore, in the following we will focus on the scale-specific
climate network with the connectivity given by the CCWT-based MIR
estimate with the wavelet coherence summation restricted to the
wavelet scales corresponding to the periods 7--8 years
(SSCN(7--8yr) thereafter). Again we consider the binary network
with $\varrho = 0.005$. The AWC of the SSCN(7--8yr) is mapped in
Fig.~\ref{nao7-8}a.
Consistently with the observation of the 7--8yr cycle in a number
of European locations ~\cite{npgsvd,jastp1}, the SSCN(7--8yr) has
a hub in a large area in Europe, but also in Western Asia and
Greenland. A strong hub of the SSCN(7--8yr) lies in the tropical
Atlantic and also in the Pacific areas different from the ENSO
area. The 7--8yr cycle in the European SAT is connected with the
North Atlantic Oscillation \cite{jastp2,pano11}.

The North Atlantic Oscillation (NAO) is a dominant pattern of the
atmospheric circulation variability in the extratropical Northern
Hemisphere.  On the global scale, the NAO has a climate
significance that rivals the Pacific ENSO \cite{marshall2001}
since it influences
the air temperature, precipitation, occurrence of storms, wind
strength and direction in the Atlantic sector and surrounding
continents. The NAO is characterized by the NAO index (NAOI, see
http://www.cru.uea.ac.uk/cru/data/nao/ for the data and their
description). While the quasi-periodic dynamics of the ENSO  is
apparent in the SOI, the NAOI has rather a red-noise-like
character \cite{naopink}. However, sensitive detection methods
such as the Monte-Carlo singular system analysis \cite{npgsvd}
uncovered in the NAO dynamics several oscillatory components from
which the cycle with the period around 7--8 years is the most
prominent \cite{jastp1,jastp2,feliks10,pano11}. The NAO 7--8yr
oscillatory mode is phase-synchronized with related modes in the
SAT in large areas of Europe \cite{pano11} as well as with other
weather and climate-related variables in various areas through the
Earth \cite{feliks10,feliks13}.

The MIR quantifying the dependence within the time scales 7--8
years between the NAOI and the SATA time series in each grid-point
is illustrated in Fig.~\ref{nao7-8}b. We can see that all the hubs
in Fig.~\ref{nao7-8}a lie in the areas where the SATA time series
are dependent on the NAOI  in this scale, with the exception of
the Pacific tropical area between 125$^{\circ}$ -- 180$^{\circ}$W
(Fig.~\ref{nao7-8}b). For the better understanding of the topology
of the SSCN(7--8yr) we quantify the dependence between the SATA
time series in the node 150$^{\circ}$W, 5$^{\circ}$N, using the
MIR estimated within the wavelet scales related to the periods
7--8 years, see Fig.~\ref{pointNAO78}a. Apparently this hub (the
Pacific tropical area between 125$^{\circ}$ -- 180$^{\circ}$W) is
connected just with other areas in the Pacific Ocean and
disconnected from the rest of the SSCN(7--8yr).

Using the same scale-specific connectivity, we can see that the
node at 50$^{\circ}$W, 7.5$^{\circ}$N in the tropical Atlantic hub
is connected to other hubs of the SSCN(7--8yr), but not to the hub
in Europe, see Fig.~\ref{pointNAO78}b. (And, of course, the
Pacific tropical hub is not connected to any other hub.) The map
of the scale-specific connectivity of a node in the European hub
(the node 22.5$^{\circ}$E, 60$^{\circ}$N lying close to the SW
Finland Baltic coast, see Fig.~\ref{pointNAOCNAO78}a) confirms the
disconnection between the tropical Atlantic and the European hubs,
however, the latter is connected to many areas all over the world.

The above-mentioned phase synchrony between the 7--8yr oscillatory
mode in the NAO and in the European SAT time series evokes the
hypothesis that, at least a part of, the connectivity in the
SSCN(7--8yr) is induced by the NAO and its world-wide influence.
In order to test this hypothesis we construct a version of the
scale-specific SSCN(7--8yr) in which, however, the connectivity is
given by the scale-specific MIR conditioned on the NAO index. In
particular, for each pair of nodes the two SATA time series and
the NAOI time series are used to construct the $3\times 3$ wavelet
coherence matrix $\Gamma^W (f)$ which is then inverted and the MIR
conditioned on the NAOI is evaluated according to Eqs.
(\ref{conwavcoh}) and (\ref{condmirg}). This measure quantifies
the scale-specific dependence between the two SATA series without
a possible influence of the NAO 7--8yr oscillatory mode. The AWC
map for this conditional SSCN(7--8yr) in
Fig.~\ref{pointNAOCNAO78}b shows that the hub in Europe and W Asia
disappears. It means that the scale 7--8yr-specific mutual
connectivity of the SATA time series in the areas in Europe and W
Asia and their connections to other areas in the world (see
Fig.~\ref{pointNAOCNAO78}a) is induced by the NAO. It is
interesting that the hub in the tropical Atlantic survived the
conditioning on the NAO, since Feliks et al.
\cite{feliks10,feliks13} track the NAO 7--8yr oscillatory mode to
 an oscillation of a similar period in the position and strength
of the Gulf Stream's sea surface temperature front in the North
Atlantic. The position of the tropical Atlantic hub coincides with
the sink region and a warm loop of the Gulf stream. So it seems
that the tropical Atlantic area plays a role in the the emergence
of the 7--8yr oscillations in nonlinear atmosphere-ocean
interactions in the Northern Atlantic, in particular in the
dynamics of the NAO. Then the NAO induces this oscillatory mode in
temperature variability in large areas in Europe, Asia as well as
in other regions in the world (Fig.~\ref{nao7-8}b). These
observations concur with some findings of  Feliks et al.
\cite{feliks13} and need further study and understanding. Detailed
insight into the related atmospheric circulation phenomena is,
however, out of the scope of this paper. Here we wanted to
demonstrate the potential of the MIR estimated using the CCWT
which gives the possibility to study either the total, or
scale-specific or conditional connectivity in networks of
interacting dynamical systems or spatio-temporal phenomena in a
discrete approximation within the complex networks paradigm.

\section{Conclusion}

Using a simple example of the autoregressive process we have
demonstrated how increasing dynamical memory, reflected, e.g., in
stronger autocorrelations, leads to increasing bias in estimating
dependence measures such as the absolute value of the correlation
coefficient. Similar behavior can be observed also in estimates of
the mutual information \cite{testpap1}, or the mean phase
coherence \cite{PhysRevE.73.065201}. We have observed how this
phenomenon can bias the connectivity in climate networks since the
time evolution of the air temperature anomalies, recorded in
different geographical areas, have different dynamics. Also in
other research fields where interaction/functional networks are
inferred from experimental time series this problem can influence
the results and skew their interpretation. For instance, many
studies of EEG functional networks reported a changed network
connectivity in different conscious states, however, changed EEG
dynamics had been reported earlier in similar experimental
conditions. The mutual information rate can be the dependence
measure which can help to distinguish changes in connectivity and
long-range synchrony from changes in the dynamics of network
nodes. Also other authors \cite{baptistaMIR,blancMIR} propose the
MIR as an association measure suitable for inferring interaction
networks from multivariate time series generated by coupled
dynamical systems. Blanc et al.~\cite{blancMIR} stress the
independence of the MIR of the time lag which can occur between
the time evolutions of two interacting systems of processes. This
property might be particularly important considering the
observation of Martin et al.~\cite{martindavidsen} regarding the
construction of the climate networks from daily air temperature
and geopotential height data. Inference of time lags in which the
maximum cross-correlation occurs in unreliable and can lead to
physically unrealistic large lags and even to the inclusion of
non-existing links to the network.

In this paper we have proposed a computationally accessible
algorithm based on the MIR of Gaussian processes, adapted by using
the wavelet transform.  We have demonstrated that this algorithm
can be effective for nonlinear, nonstationary and multiscale
processes. Using the examples of the climate networks we have
presented the ability of the scale-specific and conditional MIR to
attribute different hubs of the climate network to different
atmospheric circulation phenomena. We believe that the introduced
approach can help in further understanding of complex systems and
their dynamics which can be observed and recorded in the form of
multivariate time series.

\begin{acknowledgments}
The author would like to thank Professor A. A. Tsonis for many
inspiring discussions and the kind invitation to the workshop on
nonlinear dynamics in geosciences.

This study was supported by the Ministry of Education, Youth and
Sports of the Czech Republic within the Program KONTAKT II,
Project No. LH14001, and in its initial stage by the Czech Science
Foundation, Project No.~P103/11/J068.
\end{acknowledgments}


\begin{thebibliography}{106}%
\makeatletter
\providecommand \@ifxundefined [1]{%
 \@ifx{#1\undefined}
}%
\providecommand \@ifnum [1]{%
 \ifnum #1\expandafter \@firstoftwo
 \else \expandafter \@secondoftwo
 \fi
}%
\providecommand \@ifx [1]{%
 \ifx #1\expandafter \@firstoftwo
 \else \expandafter \@secondoftwo
 \fi
}%
\providecommand \natexlab [1]{#1}%
\providecommand \enquote  [1]{``#1''}%
\providecommand \bibnamefont  [1]{#1}%
\providecommand \bibfnamefont [1]{#1}%
\providecommand \citenamefont [1]{#1}%
\providecommand \href@noop [0]{\@secondoftwo}%
\providecommand \href [0]{\begingroup \@sanitize@url \@href}%
\providecommand \@href[1]{\@@startlink{#1}\@@href}%
\providecommand \@@href[1]{\endgroup#1\@@endlink}%
\providecommand \@sanitize@url [0]{\catcode `\\12\catcode
`\$12\catcode
  `\&12\catcode `\#12\catcode `\^12\catcode `\_12\catcode `\%12\relax}%
\providecommand \@@startlink[1]{}%
\providecommand \@@endlink[0]{}%
\providecommand \url  [0]{\begingroup\@sanitize@url \@url }%
\providecommand \@url [1]{\endgroup\@href {#1}{\urlprefix }}%
\providecommand \urlprefix  [0]{URL }%
\providecommand \Eprint [0]{\href }%
\providecommand \doibase [0]{http://dx.doi.org/}%
\providecommand \selectlanguage [0]{\@gobble}%
\providecommand \bibinfo  [0]{\@secondoftwo}%
\providecommand \bibfield  [0]{\@secondoftwo}%
\providecommand \translation [1]{[#1]}%
\providecommand \BibitemOpen [0]{}%
\providecommand \bibitemStop [0]{}%
\providecommand \bibitemNoStop [0]{.\EOS\space}%
\providecommand \EOS [0]{\spacefactor3000\relax}%
\providecommand \BibitemShut  [1]{\csname bibitem#1\endcsname}%
\let\auto@bib@innerbib\@empty
\bibitem [{\citenamefont {Soler}(2006)}]{creat}%
  \BibitemOpen
  \bibfield  {author} {\bibinfo {author} {\bibfnamefont {J.~M.}\ \bibnamefont
  {Soler}},\ }\href@noop {} {\bibfield  {journal} {\bibinfo  {journal}
  {arXiv:physics/0608006v1 [physics.soc-ph]}\ } (\bibinfo {year}
  {2006})}\BibitemShut {NoStop}%
\bibitem [{\citenamefont {Anderson}(1972)}]{more}%
  \BibitemOpen
  \bibfield  {author} {\bibinfo {author} {\bibfnamefont {P.~W.}\ \bibnamefont
  {Anderson}},\ }\href {\doibase 10.2307/1734697} {\bibfield  {journal}
  {\bibinfo  {journal} {Science}\ }\textbf {\bibinfo {volume} {177}},\ \bibinfo
  {pages} {393} (\bibinfo {year} {1972})}\BibitemShut {NoStop}%
\bibitem [{\citenamefont {Albert}\ and\ \citenamefont
  {Barab\'asi}(2002)}]{RevModPhys.74.47}%
  \BibitemOpen
  \bibfield  {author} {\bibinfo {author} {\bibfnamefont {R.}~\bibnamefont
  {Albert}}\ and\ \bibinfo {author} {\bibfnamefont {A.-L.}\ \bibnamefont
  {Barab\'asi}},\ }\href {\doibase 10.1103/RevModPhys.74.47} {\bibfield
  {journal} {\bibinfo  {journal} {Rev. Mod. Phys.}\ }\textbf {\bibinfo {volume}
  {74}},\ \bibinfo {pages} {47} (\bibinfo {year} {2002})}\BibitemShut {NoStop}%
\bibitem [{\citenamefont {Boccaletti}\ \emph {et~al.}(2006)\citenamefont
  {Boccaletti}, \citenamefont {Latora}, \citenamefont {Moreno}, \citenamefont
  {Chavez},\ and\ \citenamefont {Hwang}}]{boccaPR}%
  \BibitemOpen
  \bibfield  {author} {\bibinfo {author} {\bibfnamefont {S.}~\bibnamefont
  {Boccaletti}}, \bibinfo {author} {\bibfnamefont {V.}~\bibnamefont {Latora}},
  \bibinfo {author} {\bibfnamefont {Y.}~\bibnamefont {Moreno}}, \bibinfo
  {author} {\bibfnamefont {M.}~\bibnamefont {Chavez}}, \ and\ \bibinfo {author}
  {\bibfnamefont {D.-U.}\ \bibnamefont {Hwang}},\ }\href {\doibase
  10.1016/j.physrep.2005.10.009} {\bibfield  {journal} {\bibinfo  {journal}
  {Phys. Rep.}\ }\textbf {\bibinfo {volume} {424}},\ \bibinfo {pages} {175 }
  (\bibinfo {year} {2006})}\BibitemShut {NoStop}%
\bibitem [{\citenamefont {Newman}\ \emph {et~al.}(2006)\citenamefont {Newman},
  \citenamefont {Barab{\'a}si},\ and\ \citenamefont
  {Watts}}]{newman2006structure}%
  \BibitemOpen
  \bibfield  {author} {\bibinfo {author} {\bibfnamefont {M.}~\bibnamefont
  {Newman}}, \bibinfo {author} {\bibfnamefont {A.-L.}\ \bibnamefont
  {Barab{\'a}si}}, \ and\ \bibinfo {author} {\bibfnamefont {D.~J.}\
  \bibnamefont {Watts}},\ }\href@noop {} {\emph {\bibinfo {title} {The
  structure and dynamics of networks}}}\ (\bibinfo  {publisher} {Princeton
  University Press},\ \bibinfo {year} {2006})\BibitemShut {NoStop}%
\bibitem [{\citenamefont {Havlin}\ \emph {et~al.}(2012)\citenamefont {Havlin},
  \citenamefont {Kenett}, \citenamefont {Ben-Jacob}, \citenamefont {Bunde},
  \citenamefont {Cohen}, \citenamefont {Hermann}, \citenamefont {Kantelhardt},
  \citenamefont {Kertész}, \citenamefont {Kirkpatrick}, \citenamefont
  {Kurths}, \citenamefont {Portugali},\ and\ \citenamefont
  {Solomon}}]{havlin12}%
  \BibitemOpen
  \bibfield  {author} {\bibinfo {author} {\bibfnamefont {S.}~\bibnamefont
  {Havlin}}, \bibinfo {author} {\bibfnamefont {D.}~\bibnamefont {Kenett}},
  \bibinfo {author} {\bibfnamefont {E.}~\bibnamefont {Ben-Jacob}}, \bibinfo
  {author} {\bibfnamefont {A.}~\bibnamefont {Bunde}}, \bibinfo {author}
  {\bibfnamefont {R.}~\bibnamefont {Cohen}}, \bibinfo {author} {\bibfnamefont
  {H.}~\bibnamefont {Hermann}}, \bibinfo {author} {\bibfnamefont
  {J.}~\bibnamefont {Kantelhardt}}, \bibinfo {author} {\bibfnamefont
  {J.}~\bibnamefont {Kertész}}, \bibinfo {author} {\bibfnamefont
  {S.}~\bibnamefont {Kirkpatrick}}, \bibinfo {author} {\bibfnamefont
  {J.}~\bibnamefont {Kurths}}, \bibinfo {author} {\bibfnamefont
  {J.}~\bibnamefont {Portugali}}, \ and\ \bibinfo {author} {\bibfnamefont
  {S.}~\bibnamefont {Solomon}},\ }\href {\doibase 10.1140/epjst/e2012-01695-x}
  {\bibfield  {journal} {\bibinfo  {journal} {Eur. Phys. J. Spec. Top.}\
  }\textbf {\bibinfo {volume} {214}},\ \bibinfo {pages} {273} (\bibinfo {year}
  {2012})}\BibitemShut {NoStop}%
\bibitem [{\citenamefont {Bialonski}\ \emph {et~al.}(2010)\citenamefont
  {Bialonski}, \citenamefont {Horstmann},\ and\ \citenamefont
  {Lehnertz}}]{Bialonski2010}%
  \BibitemOpen
  \bibfield  {author} {\bibinfo {author} {\bibfnamefont {S.}~\bibnamefont
  {Bialonski}}, \bibinfo {author} {\bibfnamefont {M.-T.}\ \bibnamefont
  {Horstmann}}, \ and\ \bibinfo {author} {\bibfnamefont {K.}~\bibnamefont
  {Lehnertz}},\ }\href {\doibase http://dx.doi.org/10.1063/1.3360561}
  {\bibfield  {journal} {\bibinfo  {journal} {Chaos}\ }\textbf {\bibinfo
  {volume} {20}},\ \bibinfo {eid} {013134} (\bibinfo {year}
  {2010})}\BibitemShut {NoStop}%
\bibitem [{\citenamefont {Friston}(1994)}]{friston94}%
  \BibitemOpen
  \bibfield  {author} {\bibinfo {author} {\bibfnamefont {K.~J.}\ \bibnamefont
  {Friston}},\ }\href {\doibase 10.1002/hbm.460020107} {\bibfield  {journal}
  {\bibinfo  {journal} {Human Brain Mapping}\ }\textbf {\bibinfo {volume}
  {2}},\ \bibinfo {pages} {56} (\bibinfo {year} {1994})}\BibitemShut {NoStop}%
\bibitem [{\citenamefont {Bullmore}\ and\ \citenamefont
  {Sporns}(2009)}]{Bullmore2009}%
  \BibitemOpen
  \bibfield  {author} {\bibinfo {author} {\bibfnamefont {E.}~\bibnamefont
  {Bullmore}}\ and\ \bibinfo {author} {\bibfnamefont {O.}~\bibnamefont
  {Sporns}},\ }\href {\doibase 10.1038/nrn2575} {\bibfield  {journal} {\bibinfo
   {journal} {Nat. Rev. Neurosci.}\ }\textbf {\bibinfo {volume} {10}},\
  \bibinfo {pages} {186} (\bibinfo {year} {2009})}\BibitemShut {NoStop}%
\bibitem [{\citenamefont {Achard}\ \emph {et~al.}(2006)\citenamefont {Achard},
  \citenamefont {Salvador}, \citenamefont {Whitcher}, \citenamefont
  {Suckling},\ and\ \citenamefont {Bullmore}}]{achard06}%
  \BibitemOpen
  \bibfield  {author} {\bibinfo {author} {\bibfnamefont {S.}~\bibnamefont
  {Achard}}, \bibinfo {author} {\bibfnamefont {R.}~\bibnamefont {Salvador}},
  \bibinfo {author} {\bibfnamefont {B.}~\bibnamefont {Whitcher}}, \bibinfo
  {author} {\bibfnamefont {J.}~\bibnamefont {Suckling}}, \ and\ \bibinfo
  {author} {\bibfnamefont {E.}~\bibnamefont {Bullmore}},\ }\href {\doibase
  10.1523/JNEUROSCI.3874-05.2006} {\bibfield  {journal} {\bibinfo  {journal}
  {J. Neurosci.}\ }\textbf {\bibinfo {volume} {26}},\ \bibinfo {pages} {63}
  (\bibinfo {year} {2006})}\BibitemShut {NoStop}%
\bibitem [{\citenamefont {Reijneveld}\ \emph {et~al.}(2007)\citenamefont
  {Reijneveld}, \citenamefont {Ponten}, \citenamefont {Berendse},\ and\
  \citenamefont {Stam}}]{stam07}%
  \BibitemOpen
  \bibfield  {author} {\bibinfo {author} {\bibfnamefont {J.~C.}\ \bibnamefont
  {Reijneveld}}, \bibinfo {author} {\bibfnamefont {S.~C.}\ \bibnamefont
  {Ponten}}, \bibinfo {author} {\bibfnamefont {H.~W.}\ \bibnamefont
  {Berendse}}, \ and\ \bibinfo {author} {\bibfnamefont {C.~J.}\ \bibnamefont
  {Stam}},\ }\href {\doibase 10.1016/j.clinph.2007.08.010} {\bibfield
  {journal} {\bibinfo  {journal} {Clin. Neurophysiol.}\ }\textbf {\bibinfo
  {volume} {118}},\ \bibinfo {pages} {2317} (\bibinfo {year}
  {2007})}\BibitemShut {NoStop}%
\bibitem [{\citenamefont {Tsonis}\ and\ \citenamefont
  {Roebber}(2004)}]{tsonis2004architecture}%
  \BibitemOpen
  \bibfield  {author} {\bibinfo {author} {\bibfnamefont {A.}~\bibnamefont
  {Tsonis}}\ and\ \bibinfo {author} {\bibfnamefont {P.}~\bibnamefont
  {Roebber}},\ }\href {\doibase 10.1016/j.physa.2003.10.045} {\bibfield
  {journal} {\bibinfo  {journal} {Physica A}\ }\textbf {\bibinfo {volume}
  {333}},\ \bibinfo {pages} {497} (\bibinfo {year} {2004})}\BibitemShut
  {NoStop}%
\bibitem [{\citenamefont {Tsonis}\ \emph {et~al.}(2006)\citenamefont {Tsonis},
  \citenamefont {Swanson},\ and\ \citenamefont {Roebber}}]{tsonis2006}%
  \BibitemOpen
  \bibfield  {author} {\bibinfo {author} {\bibfnamefont {A.}~\bibnamefont
  {Tsonis}}, \bibinfo {author} {\bibfnamefont {K.}~\bibnamefont {Swanson}}, \
  and\ \bibinfo {author} {\bibfnamefont {P.}~\bibnamefont {Roebber}},\ }\href
  {\doibase 10.1175/BAMS-87-5-585} {\bibfield  {journal} {\bibinfo  {journal}
  {Bull. Amer. Meteorol. Soc.}\ }\textbf {\bibinfo {volume} {87}},\ \bibinfo
  {pages} {585} (\bibinfo {year} {2006})}\BibitemShut {NoStop}%
\bibitem [{\citenamefont {Yamasaki}\ \emph {et~al.}(2008)\citenamefont
  {Yamasaki}, \citenamefont {Gozolchiani},\ and\ \citenamefont
  {Havlin}}]{yamaninoprl}%
  \BibitemOpen
  \bibfield  {author} {\bibinfo {author} {\bibfnamefont {K.}~\bibnamefont
  {Yamasaki}}, \bibinfo {author} {\bibfnamefont {A.}~\bibnamefont
  {Gozolchiani}}, \ and\ \bibinfo {author} {\bibfnamefont {S.}~\bibnamefont
  {Havlin}},\ }\href {\doibase 10.1103/PhysRevLett.100.228501} {\bibfield
  {journal} {\bibinfo  {journal} {Phys. Rev. Lett.}\ }\textbf {\bibinfo
  {volume} {100}},\ \bibinfo {pages} {228501} (\bibinfo {year}
  {2008})}\BibitemShut {NoStop}%
\bibitem [{\citenamefont {Yamasaki}\ \emph
  {et~al.}(2009{\natexlab{a}})\citenamefont {Yamasaki}, \citenamefont
  {Gozolchiani},\ and\ \citenamefont {Havlin}}]{yamasaki09}%
  \BibitemOpen
  \bibfield  {author} {\bibinfo {author} {\bibfnamefont {K.}~\bibnamefont
  {Yamasaki}}, \bibinfo {author} {\bibfnamefont {A.}~\bibnamefont
  {Gozolchiani}}, \ and\ \bibinfo {author} {\bibfnamefont {S.}~\bibnamefont
  {Havlin}},\ }\href {http://ci.nii.ac.jp/naid/110007225713/en/} {\bibfield
  {journal} {\bibinfo  {journal} {Progr. Theor. Phys. Suppl.}\ }\textbf
  {\bibinfo {volume} {179}},\ \bibinfo {pages} {178} (\bibinfo {year}
  {2009}{\natexlab{a}})}\BibitemShut {NoStop}%
\bibitem [{\citenamefont {Donges}\ \emph
  {et~al.}(2009{\natexlab{a}})\citenamefont {Donges}, \citenamefont {Zou},
  \citenamefont {Marwan},\ and\ \citenamefont {Kurths}}]{donges2009}%
  \BibitemOpen
  \bibfield  {author} {\bibinfo {author} {\bibfnamefont {J.}~\bibnamefont
  {Donges}}, \bibinfo {author} {\bibfnamefont {Y.}~\bibnamefont {Zou}},
  \bibinfo {author} {\bibfnamefont {N.}~\bibnamefont {Marwan}}, \ and\ \bibinfo
  {author} {\bibfnamefont {J.}~\bibnamefont {Kurths}},\ }\href {\doibase
  10.1140/epjst/e2009-01098-2} {\bibfield  {journal} {\bibinfo  {journal} {Eur.
  Phys. J.-Spec. Top.}\ }\textbf {\bibinfo {volume} {174}},\ \bibinfo {pages}
  {157} (\bibinfo {year} {2009}{\natexlab{a}})}\BibitemShut {NoStop}%
\bibitem [{\citenamefont {Donges}\ \emph
  {et~al.}(2009{\natexlab{b}})\citenamefont {Donges}, \citenamefont {Zou},
  \citenamefont {Marwan},\ and\ \citenamefont {Kurths}}]{donges2009backbone}%
  \BibitemOpen
  \bibfield  {author} {\bibinfo {author} {\bibfnamefont {J.}~\bibnamefont
  {Donges}}, \bibinfo {author} {\bibfnamefont {Y.}~\bibnamefont {Zou}},
  \bibinfo {author} {\bibfnamefont {N.}~\bibnamefont {Marwan}}, \ and\ \bibinfo
  {author} {\bibfnamefont {J.}~\bibnamefont {Kurths}},\ }\href {\doibase
  10.1209/0295-5075/87/48007} {\bibfield  {journal} {\bibinfo  {journal}
  {Europhys. Lett.}\ }\textbf {\bibinfo {volume} {87}},\ \bibinfo {pages}
  {48007} (\bibinfo {year} {2009}{\natexlab{b}})}\BibitemShut {NoStop}%
\bibitem [{\citenamefont {Steinhaeuser}\ \emph {et~al.}(2012)\citenamefont
  {Steinhaeuser}, \citenamefont {Ganguly},\ and\ \citenamefont
  {Chawla}}]{gangclidym}%
  \BibitemOpen
  \bibfield  {author} {\bibinfo {author} {\bibfnamefont {K.}~\bibnamefont
  {Steinhaeuser}}, \bibinfo {author} {\bibfnamefont {A.}~\bibnamefont
  {Ganguly}}, \ and\ \bibinfo {author} {\bibfnamefont {N.}~\bibnamefont
  {Chawla}},\ }\href {\doibase 10.1007/s00382-011-1135-9} {\bibfield  {journal}
  {\bibinfo  {journal} {Clim. Dyn.}\ }\textbf {\bibinfo {volume} {39}},\
  \bibinfo {pages} {889} (\bibinfo {year} {2012})}\BibitemShut {NoStop}%
\bibitem [{\citenamefont {Schweitzer}\ \emph {et~al.}(2009)\citenamefont
  {Schweitzer}, \citenamefont {Fagiolo}, \citenamefont {Sornette},
  \citenamefont {Vega-Redondo}, \citenamefont {Vespignani},\ and\ \citenamefont
  {White}}]{economicnet}%
  \BibitemOpen
  \bibfield  {author} {\bibinfo {author} {\bibfnamefont {F.}~\bibnamefont
  {Schweitzer}}, \bibinfo {author} {\bibfnamefont {G.}~\bibnamefont {Fagiolo}},
  \bibinfo {author} {\bibfnamefont {D.}~\bibnamefont {Sornette}}, \bibinfo
  {author} {\bibfnamefont {F.}~\bibnamefont {Vega-Redondo}}, \bibinfo {author}
  {\bibfnamefont {A.}~\bibnamefont {Vespignani}}, \ and\ \bibinfo {author}
  {\bibfnamefont {D.}~\bibnamefont {White}},\ }\href {\doibase
  10.1126/science.1173644} {\bibfield  {journal} {\bibinfo  {journal}
  {Science}\ }\textbf {\bibinfo {volume} {325}},\ \bibinfo {pages} {422}
  (\bibinfo {year} {2009})}\BibitemShut {NoStop}%
\bibitem [{\citenamefont {Onnela}\ \emph {et~al.}(2004)\citenamefont {Onnela},
  \citenamefont {Kaski},\ and\ \citenamefont {Kert{\'e}sz}}]{correlfinnet}%
  \BibitemOpen
  \bibfield  {author} {\bibinfo {author} {\bibfnamefont {J.}~\bibnamefont
  {Onnela}}, \bibinfo {author} {\bibfnamefont {K.}~\bibnamefont {Kaski}}, \
  and\ \bibinfo {author} {\bibfnamefont {J.}~\bibnamefont {Kert{\'e}sz}},\
  }\href {\doibase 10.1140/epjb/e2004-00128-7} {\bibfield  {journal} {\bibinfo
  {journal} {Eur. Phys. J. B}\ }\textbf {\bibinfo {volume} {38}},\ \bibinfo
  {pages} {353} (\bibinfo {year} {2004})}\BibitemShut {NoStop}%
\bibitem [{\citenamefont {Kramer}\ \emph {et~al.}(2009)\citenamefont {Kramer},
  \citenamefont {Eden}, \citenamefont {Cash},\ and\ \citenamefont
  {Kolaczyk}}]{kramerPRE}%
  \BibitemOpen
  \bibfield  {author} {\bibinfo {author} {\bibfnamefont {M.~A.}\ \bibnamefont
  {Kramer}}, \bibinfo {author} {\bibfnamefont {U.~T.}\ \bibnamefont {Eden}},
  \bibinfo {author} {\bibfnamefont {S.~S.}\ \bibnamefont {Cash}}, \ and\
  \bibinfo {author} {\bibfnamefont {E.~D.}\ \bibnamefont {Kolaczyk}},\ }\href
  {\doibase 10.1103/PhysRevE.79.061916} {\bibfield  {journal} {\bibinfo
  {journal} {Phys. Rev. E}\ }\textbf {\bibinfo {volume} {79}},\ \bibinfo
  {pages} {061916} (\bibinfo {year} {2009})}\BibitemShut {NoStop}%
\bibitem [{\citenamefont {Palu\v{s}}\ \emph {et~al.}(2011)\citenamefont
  {Palu\v{s}}, \citenamefont {Hartman}, \citenamefont {Hlinka},\ and\
  \citenamefont {Vejmelka}}]{npgnet11}%
  \BibitemOpen
  \bibfield  {author} {\bibinfo {author} {\bibfnamefont {M.}~\bibnamefont
  {Palu\v{s}}}, \bibinfo {author} {\bibfnamefont {D.}~\bibnamefont {Hartman}},
  \bibinfo {author} {\bibfnamefont {J.}~\bibnamefont {Hlinka}}, \ and\ \bibinfo
  {author} {\bibfnamefont {M.}~\bibnamefont {Vejmelka}},\ }\href {\doibase
  10.5194/npg-18-751-2011} {\bibfield  {journal} {\bibinfo  {journal} {Nonlin.
  Processes Geophys.}\ }\textbf {\bibinfo {volume} {18}},\ \bibinfo {pages}
  {751} (\bibinfo {year} {2011})}\BibitemShut {NoStop}%
\bibitem [{\citenamefont {Bialonski}\ \emph {et~al.}(2011)\citenamefont
  {Bialonski}, \citenamefont {Wendler},\ and\ \citenamefont
  {Lehnertz}}]{Bialonski2011}%
  \BibitemOpen
  \bibfield  {author} {\bibinfo {author} {\bibfnamefont {S.}~\bibnamefont
  {Bialonski}}, \bibinfo {author} {\bibfnamefont {M.}~\bibnamefont {Wendler}},
  \ and\ \bibinfo {author} {\bibfnamefont {K.}~\bibnamefont {Lehnertz}},\
  }\href {\doibase 10.1371/journal.pone.0022826} {\bibfield  {journal}
  {\bibinfo  {journal} {PLoS One}\ }\textbf {\bibinfo {volume} {6}} (\bibinfo
  {year} {2011}),\ 10.1371/journal.pone.0022826}\BibitemShut {NoStop}%
\bibitem [{\citenamefont {Hlinka}\ \emph {et~al.}(2012)\citenamefont {Hlinka},
  \citenamefont {Hartman},\ and\ \citenamefont {Palu\v{s}}}]{hlinkaswn}%
  \BibitemOpen
  \bibfield  {author} {\bibinfo {author} {\bibfnamefont {J.}~\bibnamefont
  {Hlinka}}, \bibinfo {author} {\bibfnamefont {D.}~\bibnamefont {Hartman}}, \
  and\ \bibinfo {author} {\bibfnamefont {M.}~\bibnamefont {Palu\v{s}}},\ }\href
  {\doibase 10.1063/1.4732541} {\bibfield  {journal} {\bibinfo  {journal}
  {Chaos}\ }\textbf {\bibinfo {volume} {22}},\ \bibinfo {eid} {033107}
  (\bibinfo {year} {2012})}\BibitemShut {NoStop}%
\bibitem [{\citenamefont {Zalesky}\ \emph {et~al.}(2012)\citenamefont
  {Zalesky}, \citenamefont {Fornito},\ and\ \citenamefont
  {Bullmore}}]{Zalesky20122096}%
  \BibitemOpen
  \bibfield  {author} {\bibinfo {author} {\bibfnamefont {A.}~\bibnamefont
  {Zalesky}}, \bibinfo {author} {\bibfnamefont {A.}~\bibnamefont {Fornito}}, \
  and\ \bibinfo {author} {\bibfnamefont {E.}~\bibnamefont {Bullmore}},\ }\href
  {\doibase http://dx.doi.org/10.1016/j.neuroimage.2012.02.001} {\bibfield
  {journal} {\bibinfo  {journal} {NeuroImage}\ }\textbf {\bibinfo {volume}
  {60}},\ \bibinfo {pages} {2096 } (\bibinfo {year} {2012})}\BibitemShut
  {NoStop}%
\bibitem [{\citenamefont {Hlav\'{a}\v{c}kov\'a-Schindler}\ \emph
  {et~al.}(2007)\citenamefont {Hlav\'{a}\v{c}kov\'a-Schindler}, \citenamefont
  {Palu\v{s}}, \citenamefont {Vejmelka},\ and\ \citenamefont
  {Bhattacharya}}]{katkaPR}%
  \BibitemOpen
  \bibfield  {author} {\bibinfo {author} {\bibfnamefont {K.}~\bibnamefont
  {Hlav\'{a}\v{c}kov\'a-Schindler}}, \bibinfo {author} {\bibfnamefont
  {M.}~\bibnamefont {Palu\v{s}}}, \bibinfo {author} {\bibfnamefont
  {M.}~\bibnamefont {Vejmelka}}, \ and\ \bibinfo {author} {\bibfnamefont
  {J.}~\bibnamefont {Bhattacharya}},\ }\href {\doibase DOI
  10.1016/j.physrep.2006.12.004} {\bibfield  {journal} {\bibinfo  {journal}
  {Phys. Rep.}\ }\textbf {\bibinfo {volume} {441}},\ \bibinfo {pages} {1}
  (\bibinfo {year} {2007})}\BibitemShut {NoStop}%
\bibitem [{\citenamefont {Palu\v{s}}\ \emph {et~al.}(1993)\citenamefont
  {Palu\v{s}}, \citenamefont {Albrecht},\ and\ \citenamefont
  {Dvo\v{r}\'ak}}]{rd0}%
  \BibitemOpen
  \bibfield  {author} {\bibinfo {author} {\bibfnamefont {M.}~\bibnamefont
  {Palu\v{s}}}, \bibinfo {author} {\bibfnamefont {V.}~\bibnamefont {Albrecht}},
  \ and\ \bibinfo {author} {\bibfnamefont {I.}~\bibnamefont {Dvo\v{r}\'ak}},\
  }\href@noop {} {\bibfield  {journal} {\bibinfo  {journal} {Phys. Lett. A}\
  }\textbf {\bibinfo {volume} {175}},\ \bibinfo {pages} {203} (\bibinfo {year}
  {1993})}\BibitemShut {NoStop}%
\bibitem [{\citenamefont {Cover}\ and\ \citenamefont {Thomas}(1991)}]{thomas}%
  \BibitemOpen
  \bibfield  {author} {\bibinfo {author} {\bibfnamefont {T.}~\bibnamefont
  {Cover}}\ and\ \bibinfo {author} {\bibfnamefont {J.}~\bibnamefont {Thomas}},\
  }\href@noop {} {\emph {\bibinfo {title} {Elements of information theory}}}\
  (\bibinfo  {publisher} {Wiley, New York},\ \bibinfo {year}
  {1991})\BibitemShut {NoStop}%
\bibitem [{\citenamefont {Petersen}(1989)}]{petersen}%
  \BibitemOpen
  \bibfield  {author} {\bibinfo {author} {\bibfnamefont {K.~E.}\ \bibnamefont
  {Petersen}},\ }\href@noop {} {\emph {\bibinfo {title} {Ergodic theory}}}\
  (\bibinfo  {publisher} {Cambridge University Press},\ \bibinfo {year}
  {1989})\BibitemShut {NoStop}%
\bibitem [{\citenamefont {Pesin}(1977)}]{pesin}%
  \BibitemOpen
  \bibfield  {author} {\bibinfo {author} {\bibfnamefont {Y.~B.}\ \bibnamefont
  {Pesin}},\ }\href {http://stacks.iop.org/0036-0279/32/i=4/a=R02} {\bibfield
  {journal} {\bibinfo  {journal} {Russian Mathematical Surveys}\ }\textbf
  {\bibinfo {volume} {32}},\ \bibinfo {pages} {55} (\bibinfo {year}
  {1977})}\BibitemShut {NoStop}%
\bibitem [{\citenamefont {Grassberger}\ and\ \citenamefont
  {Procaccia}(1983{\natexlab{a}})}]{grassberger1983KSE}%
  \BibitemOpen
  \bibfield  {author} {\bibinfo {author} {\bibfnamefont {P.}~\bibnamefont
  {Grassberger}}\ and\ \bibinfo {author} {\bibfnamefont {I.}~\bibnamefont
  {Procaccia}},\ }\href@noop {} {\bibfield  {journal} {\bibinfo  {journal}
  {Phys. Rev. A}\ }\textbf {\bibinfo {volume} {28}},\ \bibinfo {pages} {2591}
  (\bibinfo {year} {1983}{\natexlab{a}})}\BibitemShut {NoStop}%
\bibitem [{\citenamefont {Grassberger}\ and\ \citenamefont
  {Procaccia}(1983{\natexlab{b}})}]{GrassbergerPhysD}%
  \BibitemOpen
  \bibfield  {author} {\bibinfo {author} {\bibfnamefont {P.}~\bibnamefont
  {Grassberger}}\ and\ \bibinfo {author} {\bibfnamefont {I.}~\bibnamefont
  {Procaccia}},\ }\href {\doibase
  http://dx.doi.org/10.1016/0167-2789(83)90298-1} {\bibfield  {journal}
  {\bibinfo  {journal} {Physica D}\ }\textbf {\bibinfo {volume} {9}},\ \bibinfo
  {pages} {189 } (\bibinfo {year} {1983}{\natexlab{b}})}\BibitemShut {NoStop}%
\bibitem [{\citenamefont {Cohen}\ and\ \citenamefont
  {Procaccia}(1985)}]{cohen85}%
  \BibitemOpen
  \bibfield  {author} {\bibinfo {author} {\bibfnamefont {A.}~\bibnamefont
  {Cohen}}\ and\ \bibinfo {author} {\bibfnamefont {I.}~\bibnamefont
  {Procaccia}},\ }\href {\doibase 10.1103/PhysRevA.31.1872} {\bibfield
  {journal} {\bibinfo  {journal} {Phys. Rev. A}\ }\textbf {\bibinfo {volume}
  {31}},\ \bibinfo {pages} {1872} (\bibinfo {year} {1985})}\BibitemShut
  {NoStop}%
\bibitem [{\citenamefont {Schouten}\ \emph {et~al.}(1994)\citenamefont
  {Schouten}, \citenamefont {Takens},\ and\ \citenamefont {van~den
  Bleek}}]{PhysRevE.49.126}%
  \BibitemOpen
  \bibfield  {author} {\bibinfo {author} {\bibfnamefont {J.~C.}\ \bibnamefont
  {Schouten}}, \bibinfo {author} {\bibfnamefont {F.}~\bibnamefont {Takens}}, \
  and\ \bibinfo {author} {\bibfnamefont {C.~M.}\ \bibnamefont {van~den
  Bleek}},\ }\href {\doibase 10.1103/PhysRevE.49.126} {\bibfield  {journal}
  {\bibinfo  {journal} {Phys. Rev. E}\ }\textbf {\bibinfo {volume} {49}},\
  \bibinfo {pages} {126} (\bibinfo {year} {1994})}\BibitemShut {NoStop}%
\bibitem [{\citenamefont {Pawelzik}\ and\ \citenamefont
  {Schuster}(1987)}]{PhysRevA.35.481}%
  \BibitemOpen
  \bibfield  {author} {\bibinfo {author} {\bibfnamefont {K.}~\bibnamefont
  {Pawelzik}}\ and\ \bibinfo {author} {\bibfnamefont {H.~G.}\ \bibnamefont
  {Schuster}},\ }\href {\doibase 10.1103/PhysRevA.35.481} {\bibfield  {journal}
  {\bibinfo  {journal} {Phys. Rev. A}\ }\textbf {\bibinfo {volume} {35}},\
  \bibinfo {pages} {481} (\bibinfo {year} {1987})}\BibitemShut {NoStop}%
\bibitem [{\citenamefont {Fraser}(1989)}]{fraser1989information}%
  \BibitemOpen
  \bibfield  {author} {\bibinfo {author} {\bibfnamefont {A.~M.}\ \bibnamefont
  {Fraser}},\ }\href {\doibase 10.1109/18.32121} {\bibfield  {journal}
  {\bibinfo  {journal} {IEEE Trans. Inf. Theory}\ }\textbf {\bibinfo {volume}
  {35}},\ \bibinfo {pages} {245} (\bibinfo {year} {1989})}\BibitemShut
  {NoStop}%
\bibitem [{\citenamefont
  {Palu\v{s}}(1997{\natexlab{a}})}]{palus1997kolmogorov}%
  \BibitemOpen
  \bibfield  {author} {\bibinfo {author} {\bibfnamefont {M.}~\bibnamefont
  {Palu\v{s}}},\ }\href@noop {} {\bibfield  {journal} {\bibinfo  {journal}
  {Neural Network World}\ }\textbf {\bibinfo {volume} {7}},\ \bibinfo {pages}
  {269} (\bibinfo {year} {1997}{\natexlab{a}})},\ \Eprint
  {http://arxiv.org/abs/http://www.cs.cas.cz/mp/papers/rd1a.pdf}
  {http://www.cs.cas.cz/mp/papers/rd1a.pdf} \BibitemShut {NoStop}%
\bibitem [{\citenamefont {Palu\v{s}}(1996{\natexlab{a}})}]{cer}%
  \BibitemOpen
  \bibfield  {author} {\bibinfo {author} {\bibfnamefont {M.}~\bibnamefont
  {Palu\v{s}}},\ }\href@noop {} {\bibfield  {journal} {\bibinfo  {journal}
  {Physica D}\ }\textbf {\bibinfo {volume} {93}},\ \bibinfo {pages} {64}
  (\bibinfo {year} {1996}{\natexlab{a}})}\BibitemShut {NoStop}%
\bibitem [{\citenamefont {Pincus}(1991)}]{pincusPNAS}%
  \BibitemOpen
  \bibfield  {author} {\bibinfo {author} {\bibfnamefont {S.~M.}\ \bibnamefont
  {Pincus}},\ }\href {\doibase 10.1073/pnas.88.6.2297} {\bibfield  {journal}
  {\bibinfo  {journal} {Proc. Natl. Acad. Sci.}\ }\textbf {\bibinfo {volume}
  {88}},\ \bibinfo {pages} {2297} (\bibinfo {year} {1991})}\BibitemShut
  {NoStop}%
\bibitem [{\citenamefont {Baptista}\ \emph {et~al.}(2010)\citenamefont
  {Baptista}, \citenamefont {Ngamga}, \citenamefont {Pinto}, \citenamefont
  {Brito},\ and\ \citenamefont {Kurths}}]{Baptista20101135}%
  \BibitemOpen
  \bibfield  {author} {\bibinfo {author} {\bibfnamefont {M.}~\bibnamefont
  {Baptista}}, \bibinfo {author} {\bibfnamefont {E.}~\bibnamefont {Ngamga}},
  \bibinfo {author} {\bibfnamefont {P.~R.}\ \bibnamefont {Pinto}}, \bibinfo
  {author} {\bibfnamefont {M.}~\bibnamefont {Brito}}, \ and\ \bibinfo {author}
  {\bibfnamefont {J.}~\bibnamefont {Kurths}},\ }\href {\doibase
  http://dx.doi.org/10.1016/j.physleta.2009.12.057} {\bibfield  {journal}
  {\bibinfo  {journal} {Phys. Lett. A}\ }\textbf {\bibinfo {volume} {374}},\
  \bibinfo {pages} {1135 } (\bibinfo {year} {2010})}\BibitemShut {NoStop}%
\bibitem [{\citenamefont {Marwan}\ \emph {et~al.}(2007)\citenamefont {Marwan},
  \citenamefont {Romano}, \citenamefont {Thiel},\ and\ \citenamefont
  {Kurths}}]{Marwan2007237}%
  \BibitemOpen
  \bibfield  {author} {\bibinfo {author} {\bibfnamefont {N.}~\bibnamefont
  {Marwan}}, \bibinfo {author} {\bibfnamefont {M.~C.}\ \bibnamefont {Romano}},
  \bibinfo {author} {\bibfnamefont {M.}~\bibnamefont {Thiel}}, \ and\ \bibinfo
  {author} {\bibfnamefont {J.}~\bibnamefont {Kurths}},\ }\href {\doibase
  http://dx.doi.org/10.1016/j.physrep.2006.11.001} {\bibfield  {journal}
  {\bibinfo  {journal} {Phys. Rep.}\ }\textbf {\bibinfo {volume} {438}},\
  \bibinfo {pages} {237 } (\bibinfo {year} {2007})}\BibitemShut {NoStop}%
\bibitem [{\citenamefont {Bandt}\ and\ \citenamefont
  {Pompe}(2002)}]{bandt2002permutation}%
  \BibitemOpen
  \bibfield  {author} {\bibinfo {author} {\bibfnamefont {C.}~\bibnamefont
  {Bandt}}\ and\ \bibinfo {author} {\bibfnamefont {B.}~\bibnamefont {Pompe}},\
  }\href@noop {} {\bibfield  {journal} {\bibinfo  {journal} {Phys. Rev. Lett.}\
  }\textbf {\bibinfo {volume} {88}},\ \bibinfo {pages} {174102} (\bibinfo
  {year} {2002})}\BibitemShut {NoStop}%
\bibitem [{\citenamefont {Lesne}\ \emph {et~al.}(2009)\citenamefont {Lesne},
  \citenamefont {Blanc},\ and\ \citenamefont {Pezard}}]{lesne09}%
  \BibitemOpen
  \bibfield  {author} {\bibinfo {author} {\bibfnamefont {A.}~\bibnamefont
  {Lesne}}, \bibinfo {author} {\bibfnamefont {J.-L.}\ \bibnamefont {Blanc}}, \
  and\ \bibinfo {author} {\bibfnamefont {L.}~\bibnamefont {Pezard}},\ }\href
  {\doibase 10.1103/PhysRevE.79.046208} {\bibfield  {journal} {\bibinfo
  {journal} {Phys. Rev. E}\ }\textbf {\bibinfo {volume} {79}},\ \bibinfo
  {pages} {046208} (\bibinfo {year} {2009})}\BibitemShut {NoStop}%
\bibitem [{\citenamefont {Ziv}\ and\ \citenamefont {Lempel}(1978)}]{lempelziv}%
  \BibitemOpen
  \bibfield  {author} {\bibinfo {author} {\bibfnamefont {J.}~\bibnamefont
  {Ziv}}\ and\ \bibinfo {author} {\bibfnamefont {A.}~\bibnamefont {Lempel}},\
  }\href {\doibase 10.1109/TIT.1978.1055934} {\bibfield  {journal} {\bibinfo
  {journal} {IEEE Trans. Inf. Theory}\ }\textbf {\bibinfo {volume} {24}},\
  \bibinfo {pages} {530} (\bibinfo {year} {1978})}\BibitemShut {NoStop}%
\bibitem [{\citenamefont {Kennel}\ \emph {et~al.}(2005)\citenamefont {Kennel},
  \citenamefont {Shlens}, \citenamefont {Abarbanel},\ and\ \citenamefont
  {Chichilnisky}}]{kennel2005}%
  \BibitemOpen
  \bibfield  {author} {\bibinfo {author} {\bibfnamefont {M.~B.}\ \bibnamefont
  {Kennel}}, \bibinfo {author} {\bibfnamefont {J.}~\bibnamefont {Shlens}},
  \bibinfo {author} {\bibfnamefont {H.~D.}\ \bibnamefont {Abarbanel}}, \ and\
  \bibinfo {author} {\bibfnamefont {E.}~\bibnamefont {Chichilnisky}},\ }\href
  {\doibase 10.1162/0899766053723050} {\bibfield  {journal} {\bibinfo
  {journal} {Neural Comput.}\ }\textbf {\bibinfo {volume} {17}},\ \bibinfo
  {pages} {1531} (\bibinfo {year} {2005})}\BibitemShut {NoStop}%
\bibitem [{\citenamefont {Crutchfield}\ and\ \citenamefont
  {Young}(1989)}]{crutch-youngPRL}%
  \BibitemOpen
  \bibfield  {author} {\bibinfo {author} {\bibfnamefont {J.~P.}\ \bibnamefont
  {Crutchfield}}\ and\ \bibinfo {author} {\bibfnamefont {K.}~\bibnamefont
  {Young}},\ }\href {\doibase 10.1103/PhysRevLett.63.105} {\bibfield  {journal}
  {\bibinfo  {journal} {Phys. Rev. Lett.}\ }\textbf {\bibinfo {volume} {63}},\
  \bibinfo {pages} {105} (\bibinfo {year} {1989})}\BibitemShut {NoStop}%
\bibitem [{\citenamefont {Shalizi}\ and\ \citenamefont
  {Crutchfield}(2001)}]{shalcrutchcompmech}%
  \BibitemOpen
  \bibfield  {author} {\bibinfo {author} {\bibfnamefont {C.}~\bibnamefont
  {Shalizi}}\ and\ \bibinfo {author} {\bibfnamefont {J.}~\bibnamefont
  {Crutchfield}},\ }\href {\doibase 10.1023/A:1010388907793} {\bibfield
  {journal} {\bibinfo  {journal} {J. Stat. Phys.}\ }\textbf {\bibinfo {volume}
  {104}},\ \bibinfo {pages} {817} (\bibinfo {year} {2001})}\BibitemShut
  {NoStop}%
\bibitem [{\citenamefont {Haslinger}\ \emph {et~al.}(2010)\citenamefont
  {Haslinger}, \citenamefont {Klinkner},\ and\ \citenamefont
  {Shalizi}}]{haslinger2010shalizi}%
  \BibitemOpen
  \bibfield  {author} {\bibinfo {author} {\bibfnamefont {R.}~\bibnamefont
  {Haslinger}}, \bibinfo {author} {\bibfnamefont {K.~L.}\ \bibnamefont
  {Klinkner}}, \ and\ \bibinfo {author} {\bibfnamefont {C.~R.}\ \bibnamefont
  {Shalizi}},\ }\href {\doibase 10.1162/neco.2009.12-07-678} {\bibfield
  {journal} {\bibinfo  {journal} {Neural Comput.}\ }\textbf {\bibinfo {volume}
  {22}},\ \bibinfo {pages} {121} (\bibinfo {year} {2010})}\BibitemShut
  {NoStop}%
\bibitem [{\citenamefont {Pinsker}(1964)}]{pinsker}%
  \BibitemOpen
  \bibfield  {author} {\bibinfo {author} {\bibfnamefont {M.~S.}\ \bibnamefont
  {Pinsker}},\ }\href@noop {} {\emph {\bibinfo {title} {Information and
  information stability of random variables and processes}}}\ (\bibinfo
  {publisher} {Holden-Day, San Francisco},\ \bibinfo {year} {1964})\BibitemShut
  {NoStop}%
\bibitem [{\citenamefont {Palu\v{s}}(1997{\natexlab{b}})}]{gser}%
  \BibitemOpen
  \bibfield  {author} {\bibinfo {author} {\bibfnamefont {M.}~\bibnamefont
  {Palu\v{s}}},\ }\href@noop {} {\bibfield  {journal} {\bibinfo  {journal}
  {Phys. Lett. A}\ }\textbf {\bibinfo {volume} {227}},\ \bibinfo {pages} {301}
  (\bibinfo {year} {1997}{\natexlab{b}})}\BibitemShut {NoStop}%
\bibitem [{\citenamefont {Palu\v{s}}(2007)}]{contempap}%
  \BibitemOpen
  \bibfield  {author} {\bibinfo {author} {\bibfnamefont {M.}~\bibnamefont
  {Palu\v{s}}},\ }\href {\doibase 10.1080/00107510801959206} {\bibfield
  {journal} {\bibinfo  {journal} {Contemp. Phys.}\ }\textbf {\bibinfo {volume}
  {48}},\ \bibinfo {pages} {307} (\bibinfo {year} {2007})}\BibitemShut
  {NoStop}%
\bibitem [{\citenamefont {Baptista}\ \emph {et~al.}(2012)\citenamefont
  {Baptista}, \citenamefont {Rubinger}, \citenamefont {Viana}, \citenamefont
  {Sartorelli}, \citenamefont {Parlitz},\ and\ \citenamefont
  {Grebogi}}]{baptistaMIR}%
  \BibitemOpen
  \bibfield  {author} {\bibinfo {author} {\bibfnamefont {M.~S.}\ \bibnamefont
  {Baptista}}, \bibinfo {author} {\bibfnamefont {R.~M.}\ \bibnamefont
  {Rubinger}}, \bibinfo {author} {\bibfnamefont {E.~R.}\ \bibnamefont {Viana}},
  \bibinfo {author} {\bibfnamefont {J.~C.}\ \bibnamefont {Sartorelli}},
  \bibinfo {author} {\bibfnamefont {U.}~\bibnamefont {Parlitz}}, \ and\
  \bibinfo {author} {\bibfnamefont {C.}~\bibnamefont {Grebogi}},\ }\href
  {\doibase 10.1371/journal.pone.0046745} {\bibfield  {journal} {\bibinfo
  {journal} {PLoS ONE}\ }\textbf {\bibinfo {volume} {7}},\ \bibinfo {pages}
  {e46745} (\bibinfo {year} {2012})}\BibitemShut {NoStop}%
\bibitem [{\citenamefont {Shlens}\ \emph {et~al.}(2007)\citenamefont {Shlens},
  \citenamefont {Kennel}, \citenamefont {Abarbanel},\ and\ \citenamefont
  {Chichilnisky}}]{shlens2007estimating}%
  \BibitemOpen
  \bibfield  {author} {\bibinfo {author} {\bibfnamefont {J.}~\bibnamefont
  {Shlens}}, \bibinfo {author} {\bibfnamefont {M.~B.}\ \bibnamefont {Kennel}},
  \bibinfo {author} {\bibfnamefont {H.~D.}\ \bibnamefont {Abarbanel}}, \ and\
  \bibinfo {author} {\bibfnamefont {E.}~\bibnamefont {Chichilnisky}},\ }\href
  {\doibase 10.1162/neco.2007.19.7.1683} {\bibfield  {journal} {\bibinfo
  {journal} {Neural Comput.}\ }\textbf {\bibinfo {volume} {19}},\ \bibinfo
  {pages} {1683} (\bibinfo {year} {2007})}\BibitemShut {NoStop}%
\bibitem [{\citenamefont {Blanc}\ \emph {et~al.}(2011)\citenamefont {Blanc},
  \citenamefont {Pezard},\ and\ \citenamefont {Lesne}}]{blancMIR}%
  \BibitemOpen
  \bibfield  {author} {\bibinfo {author} {\bibfnamefont {J.-L.}\ \bibnamefont
  {Blanc}}, \bibinfo {author} {\bibfnamefont {L.}~\bibnamefont {Pezard}}, \
  and\ \bibinfo {author} {\bibfnamefont {A.}~\bibnamefont {Lesne}},\ }\href
  {\doibase 10.1103/PhysRevE.84.036214} {\bibfield  {journal} {\bibinfo
  {journal} {Phys. Rev. E}\ }\textbf {\bibinfo {volume} {84}},\ \bibinfo
  {pages} {036214} (\bibinfo {year} {2011})}\BibitemShut {NoStop}%
\bibitem [{\citenamefont {Jiruska}\ \emph {et~al.}(2010)\citenamefont
  {Jiruska}, \citenamefont {Csicsvari}, \citenamefont {Powell}, \citenamefont
  {Fox}, \citenamefont {Chang}, \citenamefont {Vreugdenhil}, \citenamefont
  {Li}, \citenamefont {Palus}, \citenamefont {Bujan}, \citenamefont {Dearden}
  \emph {et~al.}}]{jiruska2010}%
  \BibitemOpen
  \bibfield  {author} {\bibinfo {author} {\bibfnamefont {P.}~\bibnamefont
  {Jiruska}}, \bibinfo {author} {\bibfnamefont {J.}~\bibnamefont {Csicsvari}},
  \bibinfo {author} {\bibfnamefont {A.~D.}\ \bibnamefont {Powell}}, \bibinfo
  {author} {\bibfnamefont {J.~E.}\ \bibnamefont {Fox}}, \bibinfo {author}
  {\bibfnamefont {W.-C.}\ \bibnamefont {Chang}}, \bibinfo {author}
  {\bibfnamefont {M.}~\bibnamefont {Vreugdenhil}}, \bibinfo {author}
  {\bibfnamefont {X.}~\bibnamefont {Li}}, \bibinfo {author} {\bibfnamefont
  {M.}~\bibnamefont {Palus}}, \bibinfo {author} {\bibfnamefont {A.~F.}\
  \bibnamefont {Bujan}}, \bibinfo {author} {\bibfnamefont {R.~W.}\ \bibnamefont
  {Dearden}},  \emph {et~al.},\ }\href {\doibase
  10.1523/JNEUROSCI.0535-10.2010} {\bibfield  {journal} {\bibinfo  {journal}
  {J. Neurosci.}\ }\textbf {\bibinfo {volume} {30}},\ \bibinfo {pages} {5690}
  (\bibinfo {year} {2010})}\BibitemShut {NoStop}%
\bibitem [{\citenamefont {Casdagli}\ \emph {et~al.}(1996)\citenamefont
  {Casdagli}, \citenamefont {Iasemidis}, \citenamefont {Sackellares},
  \citenamefont {Roper}, \citenamefont {Gilmore},\ and\ \citenamefont
  {Savit}}]{Casdagli1996381}%
  \BibitemOpen
  \bibfield  {author} {\bibinfo {author} {\bibfnamefont {M.}~\bibnamefont
  {Casdagli}}, \bibinfo {author} {\bibfnamefont {L.}~\bibnamefont {Iasemidis}},
  \bibinfo {author} {\bibfnamefont {J.}~\bibnamefont {Sackellares}}, \bibinfo
  {author} {\bibfnamefont {S.}~\bibnamefont {Roper}}, \bibinfo {author}
  {\bibfnamefont {R.}~\bibnamefont {Gilmore}}, \ and\ \bibinfo {author}
  {\bibfnamefont {R.}~\bibnamefont {Savit}},\ }\href {\doibase
  http://dx.doi.org/10.1016/S0167-2789(96)00160-1} {\bibfield  {journal}
  {\bibinfo  {journal} {Physica D}\ }\textbf {\bibinfo {volume} {99}},\
  \bibinfo {pages} {381 } (\bibinfo {year} {1996})}\BibitemShut {NoStop}%
\bibitem [{\citenamefont {Palu\v{s}}\ and\ \citenamefont
  {Vejmelka}(2007)}]{testpap1}%
  \BibitemOpen
  \bibfield  {author} {\bibinfo {author} {\bibfnamefont {M.}~\bibnamefont
  {Palu\v{s}}}\ and\ \bibinfo {author} {\bibfnamefont {M.}~\bibnamefont
  {Vejmelka}},\ }\href {\doibase 10.1103/PhysRevE.75.056211} {\bibfield
  {journal} {\bibinfo  {journal} {Phys. Rev. E}\ }\textbf {\bibinfo {volume}
  {75}},\ \bibinfo {pages} {056211} (\bibinfo {year} {2007})}\BibitemShut
  {NoStop}%
\bibitem [{\citenamefont {Palu\v{s}}\ \emph {et~al.}(2001)\citenamefont
  {Palu\v{s}}, \citenamefont {Komarek}, \citenamefont {Hrncir},\ and\
  \citenamefont {Sterbova}}]{cir1pre}%
  \BibitemOpen
  \bibfield  {author} {\bibinfo {author} {\bibfnamefont {M.}~\bibnamefont
  {Palu\v{s}}}, \bibinfo {author} {\bibfnamefont {V.}~\bibnamefont {Komarek}},
  \bibinfo {author} {\bibfnamefont {Z.}~\bibnamefont {Hrncir}}, \ and\ \bibinfo
  {author} {\bibfnamefont {K.}~\bibnamefont {Sterbova}},\ }\href {\doibase
  10.1103/PhysRevE.63.046211} {\bibfield  {journal} {\bibinfo  {journal} {Phys.
  Rev. E}\ }\textbf {\bibinfo {volume} {63}},\ \bibinfo {pages} {046211}
  (\bibinfo {year} {2001})}\BibitemShut {NoStop}%
\bibitem [{\citenamefont {Hlinka}\ \emph {et~al.}(2011)\citenamefont {Hlinka},
  \citenamefont {Palu\v{s}}, \citenamefont {Vejmelka}, \citenamefont
  {Mantini},\ and\ \citenamefont {Corbetta}}]{hlinkalinfmri}%
  \BibitemOpen
  \bibfield  {author} {\bibinfo {author} {\bibfnamefont {J.}~\bibnamefont
  {Hlinka}}, \bibinfo {author} {\bibfnamefont {M.}~\bibnamefont {Palu\v{s}}},
  \bibinfo {author} {\bibfnamefont {M.}~\bibnamefont {Vejmelka}}, \bibinfo
  {author} {\bibfnamefont {D.}~\bibnamefont {Mantini}}, \ and\ \bibinfo
  {author} {\bibfnamefont {M.}~\bibnamefont {Corbetta}},\ }\href {\doibase
  http://dx.doi.org/10.1016/j.neuroimage.2010.08.042} {\bibfield  {journal}
  {\bibinfo  {journal} {NeuroImage}\ }\textbf {\bibinfo {volume} {54}},\
  \bibinfo {pages} {2218 } (\bibinfo {year} {2011})}\BibitemShut {NoStop}%
\bibitem [{\citenamefont {{Hartman}}\ \emph {et~al.}(2011)\citenamefont
  {{Hartman}}, \citenamefont {{Hlinka}}, \citenamefont {{Palu{\v s}}},
  \citenamefont {{Mantini}},\ and\ \citenamefont {{Corbetta}}}]{hartman}%
  \BibitemOpen
  \bibfield  {author} {\bibinfo {author} {\bibfnamefont {D.}~\bibnamefont
  {{Hartman}}}, \bibinfo {author} {\bibfnamefont {J.}~\bibnamefont {{Hlinka}}},
  \bibinfo {author} {\bibfnamefont {M.}~\bibnamefont {{Palu{\v s}}}}, \bibinfo
  {author} {\bibfnamefont {D.}~\bibnamefont {{Mantini}}}, \ and\ \bibinfo
  {author} {\bibfnamefont {M.}~\bibnamefont {{Corbetta}}},\ }\href {\doibase
  10.1063/1.3553181} {\bibfield  {journal} {\bibinfo  {journal} {Chaos}\
  }\textbf {\bibinfo {volume} {21}},\ \bibinfo {pages} {013119} (\bibinfo
  {year} {2011})}\BibitemShut {NoStop}%
\bibitem [{\citenamefont {Palu\v{s}}(1996{\natexlab{b}})}]{rdbc}%
  \BibitemOpen
  \bibfield  {author} {\bibinfo {author} {\bibfnamefont {M.}~\bibnamefont
  {Palu\v{s}}},\ }\href@noop {} {\bibfield  {journal} {\bibinfo  {journal}
  {Biol. Cybern.}\ }\textbf {\bibinfo {volume} {75}},\ \bibinfo {pages} {389}
  (\bibinfo {year} {1996}{\natexlab{b}})}\BibitemShut {NoStop}%
\bibitem [{\citenamefont {Theiler}\ and\ \citenamefont
  {Rapp}(1996)}]{theirapp}%
  \BibitemOpen
  \bibfield  {author} {\bibinfo {author} {\bibfnamefont {J.}~\bibnamefont
  {Theiler}}\ and\ \bibinfo {author} {\bibfnamefont {P.~E.}\ \bibnamefont
  {Rapp}},\ }\href {\doibase http://dx.doi.org/10.1016/0013-4694(95)00240-5}
  {\bibfield  {journal} {\bibinfo  {journal} {Electroencephalography and
  Clinical Neurophysiology}\ }\textbf {\bibinfo {volume} {98}},\ \bibinfo
  {pages} {213 } (\bibinfo {year} {1996})}\BibitemShut {NoStop}%
\bibitem [{\citenamefont {Matousek}\ \emph {et~al.}(1995)\citenamefont
  {Matousek}, \citenamefont {Wackermann}, \citenamefont {Palus}, \citenamefont
  {Berankova}, \citenamefont {Albrecht},\ and\ \citenamefont
  {Dvorak}}]{matousek1995global}%
  \BibitemOpen
  \bibfield  {author} {\bibinfo {author} {\bibfnamefont {M.}~\bibnamefont
  {Matousek}}, \bibinfo {author} {\bibfnamefont {J.}~\bibnamefont
  {Wackermann}}, \bibinfo {author} {\bibfnamefont {M.}~\bibnamefont {Palus}},
  \bibinfo {author} {\bibfnamefont {A.}~\bibnamefont {Berankova}}, \bibinfo
  {author} {\bibfnamefont {V.}~\bibnamefont {Albrecht}}, \ and\ \bibinfo
  {author} {\bibfnamefont {I.}~\bibnamefont {Dvorak}},\ }\href {\doibase
  10.1159/000119171} {\bibfield  {journal} {\bibinfo  {journal}
  {Neuropsychobiology}\ }\textbf {\bibinfo {volume} {31}},\ \bibinfo {pages}
  {47} (\bibinfo {year} {1995})}\BibitemShut {NoStop}%
\bibitem [{\citenamefont {Palu\v{s}}\ \emph {et~al.}(1992)\citenamefont
  {Palu\v{s}}, \citenamefont {Dvorak},\ and\ \citenamefont {David}}]{spat0}%
  \BibitemOpen
  \bibfield  {author} {\bibinfo {author} {\bibfnamefont {M.}~\bibnamefont
  {Palu\v{s}}}, \bibinfo {author} {\bibfnamefont {I.}~\bibnamefont {Dvorak}}, \
  and\ \bibinfo {author} {\bibfnamefont {I.}~\bibnamefont {David}},\
  }\href@noop {} {\bibfield  {journal} {\bibinfo  {journal} {Physica A}\
  }\textbf {\bibinfo {volume} {185}},\ \bibinfo {pages} {433} (\bibinfo {year}
  {1992})}\BibitemShut {NoStop}%
\bibitem [{\citenamefont {Nicolis}\ and\ \citenamefont
  {Nicolis}(1984)}]{nicolis84}%
  \BibitemOpen
  \bibfield  {author} {\bibinfo {author} {\bibfnamefont {C.}~\bibnamefont
  {Nicolis}}\ and\ \bibinfo {author} {\bibfnamefont {G.}~\bibnamefont
  {Nicolis}},\ }\href {\doibase 10.1038/311529a0} {\bibfield  {journal}
  {\bibinfo  {journal} {Nature}\ }\textbf {\bibinfo {volume} {311}},\ \bibinfo
  {pages} {529} (\bibinfo {year} {1984})}\BibitemShut {NoStop}%
\bibitem [{\citenamefont {Fraedrich}(1986)}]{fraedrich86}%
  \BibitemOpen
  \bibfield  {author} {\bibinfo {author} {\bibfnamefont {K.}~\bibnamefont
  {Fraedrich}},\ }\href {\doibase
  10.1175/1520-0469(1986)043<0419:ETDOWA>2.0.CO;2} {\bibfield  {journal}
  {\bibinfo  {journal} {J. Atmos. Sci.}\ }\textbf {\bibinfo {volume} {43}},\
  \bibinfo {pages} {419} (\bibinfo {year} {1986})}\BibitemShut {NoStop}%
\bibitem [{\citenamefont {{Tsonis}}\ and\ \citenamefont
  {{Elsner}}(1988)}]{tsonis88}%
  \BibitemOpen
  \bibfield  {author} {\bibinfo {author} {\bibfnamefont {A.~A.}\ \bibnamefont
  {{Tsonis}}}\ and\ \bibinfo {author} {\bibfnamefont {J.~B.}\ \bibnamefont
  {{Elsner}}},\ }\href {\doibase 10.1038/333545a0} {\bibfield  {journal}
  {\bibinfo  {journal} {Nature}\ }\textbf {\bibinfo {volume} {333}},\ \bibinfo
  {pages} {545} (\bibinfo {year} {1988})}\BibitemShut {NoStop}%
\bibitem [{\citenamefont {{Grassberger}}(1986)}]{grassberger86}%
  \BibitemOpen
  \bibfield  {author} {\bibinfo {author} {\bibfnamefont {P.}~\bibnamefont
  {{Grassberger}}},\ }\href {\doibase 10.1038/323609a0} {\bibfield  {journal}
  {\bibinfo  {journal} {Nature}\ }\textbf {\bibinfo {volume} {323}},\ \bibinfo
  {pages} {609} (\bibinfo {year} {1986})}\BibitemShut {NoStop}%
\bibitem [{\citenamefont {{Lorenz}}(1991)}]{lorenz91}%
  \BibitemOpen
  \bibfield  {author} {\bibinfo {author} {\bibfnamefont {E.~N.}\ \bibnamefont
  {{Lorenz}}},\ }\href {\doibase 10.1038/353241a0} {\bibfield  {journal}
  {\bibinfo  {journal} {Nature}\ }\textbf {\bibinfo {volume} {353}},\ \bibinfo
  {pages} {241} (\bibinfo {year} {1991})}\BibitemShut {NoStop}%
\bibitem [{\citenamefont {Palu\v{s}}\ and\ \citenamefont
  {Novotn\'{a}}(1994)}]{meteo}%
  \BibitemOpen
  \bibfield  {author} {\bibinfo {author} {\bibfnamefont {M.}~\bibnamefont
  {Palu\v{s}}}\ and\ \bibinfo {author} {\bibfnamefont {D.}~\bibnamefont
  {Novotn\'{a}}},\ }\href@noop {} {\bibfield  {journal} {\bibinfo  {journal}
  {Phys. Lett. A}\ }\textbf {\bibinfo {volume} {193}},\ \bibinfo {pages} {67}
  (\bibinfo {year} {1994})}\BibitemShut {NoStop}%
\bibitem [{\citenamefont {Hlinka}\ \emph
  {et~al.}(2013{\natexlab{a}})\citenamefont {Hlinka}, \citenamefont {Hartman},
  \citenamefont {Vejmelka}, \citenamefont {Novotn\'a},\ and\ \citenamefont
  {Palu\v{s}}}]{hlinkaclidy}%
  \BibitemOpen
  \bibfield  {author} {\bibinfo {author} {\bibfnamefont {J.}~\bibnamefont
  {Hlinka}}, \bibinfo {author} {\bibfnamefont {D.}~\bibnamefont {Hartman}},
  \bibinfo {author} {\bibfnamefont {M.}~\bibnamefont {Vejmelka}}, \bibinfo
  {author} {\bibfnamefont {D.}~\bibnamefont {Novotn\'a}}, \ and\ \bibinfo
  {author} {\bibfnamefont {M.}~\bibnamefont {Palu\v{s}}},\ }\href {\doibase
  10.1007/s00382-013-1780-2} {\bibfield  {journal} {\bibinfo  {journal} {Clim.
  Dyn.}\ } (\bibinfo {year} {2013}{\natexlab{a}}),\
  10.1007/s00382-013-1780-2}\BibitemShut {NoStop}%
\bibitem [{\citenamefont {Palu\v{s}}\ and\ \citenamefont
  {Novotn\'a}(2004)}]{npgsvd}%
  \BibitemOpen
  \bibfield  {author} {\bibinfo {author} {\bibfnamefont {M.}~\bibnamefont
  {Palu\v{s}}}\ and\ \bibinfo {author} {\bibfnamefont {D.}~\bibnamefont
  {Novotn\'a}},\ }\href@noop {} {\bibfield  {journal} {\bibinfo  {journal}
  {Nonlin. Processes Geophys.}\ }\textbf {\bibinfo {volume} {11}},\ \bibinfo
  {pages} {721} (\bibinfo {year} {2004})}\BibitemShut {NoStop}%
\bibitem [{\citenamefont {Palu\v{s}}\ and\ \citenamefont
  {Novotn\'a}(2006)}]{npgqbo}%
  \BibitemOpen
  \bibfield  {author} {\bibinfo {author} {\bibfnamefont {M.}~\bibnamefont
  {Palu\v{s}}}\ and\ \bibinfo {author} {\bibfnamefont {D.}~\bibnamefont
  {Novotn\'a}},\ }\href@noop {} {\bibfield  {journal} {\bibinfo  {journal}
  {Nonlin. Processes Geophys.}\ }\textbf {\bibinfo {volume} {13}},\ \bibinfo
  {pages} {287} (\bibinfo {year} {2006})}\BibitemShut {NoStop}%
\bibitem [{\citenamefont {Palu\v{s}}\ and\ \citenamefont
  {Novotn\'a}(2009)}]{jastp2}%
  \BibitemOpen
  \bibfield  {author} {\bibinfo {author} {\bibfnamefont {M.}~\bibnamefont
  {Palu\v{s}}}\ and\ \bibinfo {author} {\bibfnamefont {D.}~\bibnamefont
  {Novotn\'a}},\ }\href {\doibase DOI 10.1016/j.jastp.2009.03.012} {\bibfield
  {journal} {\bibinfo  {journal} {J. Atmos. Sol.-Terr. Phys.}\ }\textbf
  {\bibinfo {volume} {71}},\ \bibinfo {pages} {923} (\bibinfo {year}
  {2009})}\BibitemShut {NoStop}%
\bibitem [{\citenamefont {{Feliks}}\ \emph {et~al.}(2010)\citenamefont
  {{Feliks}}, \citenamefont {{Ghil}},\ and\ \citenamefont
  {{Robertson}}}]{feliks10}%
  \BibitemOpen
  \bibfield  {author} {\bibinfo {author} {\bibfnamefont {Y.}~\bibnamefont
  {{Feliks}}}, \bibinfo {author} {\bibfnamefont {M.}~\bibnamefont {{Ghil}}}, \
  and\ \bibinfo {author} {\bibfnamefont {A.~W.}\ \bibnamefont {{Robertson}}},\
  }\href {\doibase 10.1175/2010JCLI3181.1} {\bibfield  {journal} {\bibinfo
  {journal} {J. Climate}\ }\textbf {\bibinfo {volume} {23}},\ \bibinfo {pages}
  {4060} (\bibinfo {year} {2010})}\BibitemShut {NoStop}%
\bibitem [{\citenamefont {Palu\v{s}}\ and\ \citenamefont
  {Novotn\'a}(2011)}]{pano11}%
  \BibitemOpen
  \bibfield  {author} {\bibinfo {author} {\bibfnamefont {M.}~\bibnamefont
  {Palu\v{s}}}\ and\ \bibinfo {author} {\bibfnamefont {D.}~\bibnamefont
  {Novotn\'a}},\ }\href {\doibase 10.5194/npg-18-251-2011} {\bibfield
  {journal} {\bibinfo  {journal} {Nonlin. Processes Geophys.}\ }\textbf
  {\bibinfo {volume} {18}},\ \bibinfo {pages} {251} (\bibinfo {year}
  {2011})}\BibitemShut {NoStop}%
\bibitem [{\citenamefont {Hlinka}\ \emph
  {et~al.}(2013{\natexlab{b}})\citenamefont {Hlinka}, \citenamefont {Hartman},
  \citenamefont {Vejmelka}, \citenamefont {Runge}, \citenamefont {Marwan},
  \citenamefont {Kurths},\ and\ \citenamefont {Palu\v{s}}}]{hlinkaentro13}%
  \BibitemOpen
  \bibfield  {author} {\bibinfo {author} {\bibfnamefont {J.}~\bibnamefont
  {Hlinka}}, \bibinfo {author} {\bibfnamefont {D.}~\bibnamefont {Hartman}},
  \bibinfo {author} {\bibfnamefont {M.}~\bibnamefont {Vejmelka}}, \bibinfo
  {author} {\bibfnamefont {J.}~\bibnamefont {Runge}}, \bibinfo {author}
  {\bibfnamefont {N.}~\bibnamefont {Marwan}}, \bibinfo {author} {\bibfnamefont
  {J.}~\bibnamefont {Kurths}}, \ and\ \bibinfo {author} {\bibfnamefont
  {M.}~\bibnamefont {Palu\v{s}}},\ }\href {\doibase 10.3390/e15062023}
  {\bibfield  {journal} {\bibinfo  {journal} {Entropy}\ }\textbf {\bibinfo
  {volume} {15}},\ \bibinfo {pages} {2023} (\bibinfo {year}
  {2013}{\natexlab{b}})}\BibitemShut {NoStop}%
\bibitem [{\citenamefont {Holme}\ and\ \citenamefont
  {Saram\"{a}ki}(2012)}]{temporalnets}%
  \BibitemOpen
  \bibfield  {author} {\bibinfo {author} {\bibfnamefont {P.}~\bibnamefont
  {Holme}}\ and\ \bibinfo {author} {\bibfnamefont {J.}~\bibnamefont
  {Saram\"{a}ki}},\ }\href {\doibase
  http://dx.doi.org/10.1016/j.physrep.2012.03.001} {\bibfield  {journal}
  {\bibinfo  {journal} {Phys. Rep.}\ }\textbf {\bibinfo {volume} {519}},\
  \bibinfo {pages} {97 } (\bibinfo {year} {2012})}\BibitemShut {NoStop}%
\bibitem [{\citenamefont {Kuhnert}\ \emph {et~al.}(2010)\citenamefont
  {Kuhnert}, \citenamefont {Elger},\ and\ \citenamefont
  {Lehnertz}}]{lehnertz-evo1}%
  \BibitemOpen
  \bibfield  {author} {\bibinfo {author} {\bibfnamefont {M.-T.}\ \bibnamefont
  {Kuhnert}}, \bibinfo {author} {\bibfnamefont {C.~E.}\ \bibnamefont {Elger}},
  \ and\ \bibinfo {author} {\bibfnamefont {K.}~\bibnamefont {Lehnertz}},\
  }\href {\doibase http://dx.doi.org/10.1063/1.3504998} {\bibfield  {journal}
  {\bibinfo  {journal} {Chaos}\ }\textbf {\bibinfo {volume} {20}},\ \bibinfo
  {eid} {043126} (\bibinfo {year} {2010})}\BibitemShut {NoStop}%
\bibitem [{\citenamefont {Bialonski}\ and\ \citenamefont
  {Lehnertz}(2013)}]{lehnertz-evo2}%
  \BibitemOpen
  \bibfield  {author} {\bibinfo {author} {\bibfnamefont {S.}~\bibnamefont
  {Bialonski}}\ and\ \bibinfo {author} {\bibfnamefont {K.}~\bibnamefont
  {Lehnertz}},\ }\href {\doibase http://dx.doi.org/10.1063/1.4821915}
  {\bibfield  {journal} {\bibinfo  {journal} {Chaos}\ }\textbf {\bibinfo
  {volume} {23}},\ \bibinfo {eid} {033139} (\bibinfo {year}
  {2013})}\BibitemShut {NoStop}%
\bibitem [{\citenamefont {Lehnertz}\ \emph {et~al.}(2014)\citenamefont
  {Lehnertz}, \citenamefont {Ansmann}, \citenamefont {Bialonski}, \citenamefont
  {Dickten}, \citenamefont {Geier},\ and\ \citenamefont
  {Porz}}]{lehnertz-evo3}%
  \BibitemOpen
  \bibfield  {author} {\bibinfo {author} {\bibfnamefont {K.}~\bibnamefont
  {Lehnertz}}, \bibinfo {author} {\bibfnamefont {G.}~\bibnamefont {Ansmann}},
  \bibinfo {author} {\bibfnamefont {S.}~\bibnamefont {Bialonski}}, \bibinfo
  {author} {\bibfnamefont {H.}~\bibnamefont {Dickten}}, \bibinfo {author}
  {\bibfnamefont {C.}~\bibnamefont {Geier}}, \ and\ \bibinfo {author}
  {\bibfnamefont {S.}~\bibnamefont {Porz}},\ }\href {\doibase
  http://dx.doi.org/10.1016/j.physd.2013.06.009} {\bibfield  {journal}
  {\bibinfo  {journal} {Physica D}\ }\textbf {\bibinfo {volume} {267}},\
  \bibinfo {pages} {7 } (\bibinfo {year} {2014})}\BibitemShut {NoStop}%
\bibitem [{\citenamefont {Radebach}\ \emph {et~al.}(2013)\citenamefont
  {Radebach}, \citenamefont {Donner}, \citenamefont {Runge}, \citenamefont
  {Donges},\ and\ \citenamefont {Kurths}}]{radebachPRE}%
  \BibitemOpen
  \bibfield  {author} {\bibinfo {author} {\bibfnamefont {A.}~\bibnamefont
  {Radebach}}, \bibinfo {author} {\bibfnamefont {R.~V.}\ \bibnamefont
  {Donner}}, \bibinfo {author} {\bibfnamefont {J.}~\bibnamefont {Runge}},
  \bibinfo {author} {\bibfnamefont {J.~F.}\ \bibnamefont {Donges}}, \ and\
  \bibinfo {author} {\bibfnamefont {J.}~\bibnamefont {Kurths}},\ }\href
  {\doibase 10.1103/PhysRevE.88.052807} {\bibfield  {journal} {\bibinfo
  {journal} {Phys. Rev. E}\ }\textbf {\bibinfo {volume} {88}},\ \bibinfo
  {pages} {052807} (\bibinfo {year} {2013})}\BibitemShut {NoStop}%
\bibitem [{\citenamefont {Tsonis}\ and\ \citenamefont
  {Swanson}(2008)}]{tsonis2008prl}%
  \BibitemOpen
  \bibfield  {author} {\bibinfo {author} {\bibfnamefont {A.}~\bibnamefont
  {Tsonis}}\ and\ \bibinfo {author} {\bibfnamefont {K.}~\bibnamefont
  {Swanson}},\ }\href {\doibase 10.1103/PhysRevLett.100.228502} {\bibfield
  {journal} {\bibinfo  {journal} {Phys. Rev. Lett.}\ }\textbf {\bibinfo
  {volume} {100}},\ \bibinfo {pages} {228502} (\bibinfo {year}
  {2008})}\BibitemShut {NoStop}%
\bibitem [{\citenamefont {Torrence}\ and\ \citenamefont {Compo}(1998)}]{ccwt}%
  \BibitemOpen
  \bibfield  {author} {\bibinfo {author} {\bibfnamefont {C.}~\bibnamefont
  {Torrence}}\ and\ \bibinfo {author} {\bibfnamefont {G.~P.}\ \bibnamefont
  {Compo}},\ }\href@noop {} {\bibfield  {journal} {\bibinfo  {journal} {Bull.
  Amer. Meteor. Soc.}\ }\textbf {\bibinfo {volume} {79}},\ \bibinfo {pages}
  {61} (\bibinfo {year} {1998})}\BibitemShut {NoStop}%
\bibitem [{\citenamefont {Theiler}\ \emph {et~al.}(1992)\citenamefont
  {Theiler}, \citenamefont {Eubank}, \citenamefont {Longtin}, \citenamefont
  {Galdrikian},\ and\ \citenamefont {Farmer}}]{Theiler1992}%
  \BibitemOpen
  \bibfield  {author} {\bibinfo {author} {\bibfnamefont {J.}~\bibnamefont
  {Theiler}}, \bibinfo {author} {\bibfnamefont {S.}~\bibnamefont {Eubank}},
  \bibinfo {author} {\bibfnamefont {A.}~\bibnamefont {Longtin}}, \bibinfo
  {author} {\bibfnamefont {B.}~\bibnamefont {Galdrikian}}, \ and\ \bibinfo
  {author} {\bibfnamefont {J.~D.}\ \bibnamefont {Farmer}},\ }\href {\doibase
  http://dx.doi.org/10.1016/0167-2789(92)90102-S} {\bibfield  {journal}
  {\bibinfo  {journal} {Physica D}\ }\textbf {\bibinfo {volume} {58}},\
  \bibinfo {pages} {77 } (\bibinfo {year} {1992})}\BibitemShut {NoStop}%
\bibitem [{\citenamefont {Schelter}\ \emph {et~al.}(2006)\citenamefont
  {Schelter}, \citenamefont {Winterhalder}, \citenamefont {Dahlhaus},
  \citenamefont {Kurths},\ and\ \citenamefont {Timmer}}]{schelterpartial}%
  \BibitemOpen
  \bibfield  {author} {\bibinfo {author} {\bibfnamefont {B.}~\bibnamefont
  {Schelter}}, \bibinfo {author} {\bibfnamefont {M.}~\bibnamefont
  {Winterhalder}}, \bibinfo {author} {\bibfnamefont {R.}~\bibnamefont
  {Dahlhaus}}, \bibinfo {author} {\bibfnamefont {J.}~\bibnamefont {Kurths}}, \
  and\ \bibinfo {author} {\bibfnamefont {J.}~\bibnamefont {Timmer}},\ }\href
  {\doibase 10.1103/PhysRevLett.96.208103} {\bibfield  {journal} {\bibinfo
  {journal} {Phys. Rev. Lett.}\ }\textbf {\bibinfo {volume} {96}},\ \bibinfo
  {pages} {208103} (\bibinfo {year} {2006})}\BibitemShut {NoStop}%
\bibitem [{\citenamefont {Kalnay}\ \emph {et~al.}(1996)\citenamefont {Kalnay},
  \citenamefont {Kanamitsu}, \citenamefont {Kistler}, \citenamefont {Collins},
  \citenamefont {Deaven}, \citenamefont {Gandin}, \citenamefont {Iredell},
  \citenamefont {Saha}, \citenamefont {White}, \citenamefont {Woollen} \emph
  {et~al.}}]{ncep}%
  \BibitemOpen
  \bibfield  {author} {\bibinfo {author} {\bibfnamefont {E.}~\bibnamefont
  {Kalnay}}, \bibinfo {author} {\bibfnamefont {M.}~\bibnamefont {Kanamitsu}},
  \bibinfo {author} {\bibfnamefont {R.}~\bibnamefont {Kistler}}, \bibinfo
  {author} {\bibfnamefont {W.}~\bibnamefont {Collins}}, \bibinfo {author}
  {\bibfnamefont {D.}~\bibnamefont {Deaven}}, \bibinfo {author} {\bibfnamefont
  {L.}~\bibnamefont {Gandin}}, \bibinfo {author} {\bibfnamefont
  {M.}~\bibnamefont {Iredell}}, \bibinfo {author} {\bibfnamefont
  {S.}~\bibnamefont {Saha}}, \bibinfo {author} {\bibfnamefont {G.}~\bibnamefont
  {White}}, \bibinfo {author} {\bibfnamefont {J.}~\bibnamefont {Woollen}},
  \emph {et~al.},\ }\href {\doibase
  10.1175/1520-0477(1996)077<0437:TNYRP>2.0.CO;2} {\bibfield  {journal}
  {\bibinfo  {journal} {Bull. Am. Met. Soc.}\ }\textbf {\bibinfo {volume}
  {77}},\ \bibinfo {pages} {437} (\bibinfo {year} {1996})}\BibitemShut
  {NoStop}%
\bibitem [{\citenamefont {Gozolchiani}\ \emph {et~al.}(2008)\citenamefont
  {Gozolchiani}, \citenamefont {Yamasaki}, \citenamefont {Gazit},\ and\
  \citenamefont {Havlin}}]{yamaninoepl}%
  \BibitemOpen
  \bibfield  {author} {\bibinfo {author} {\bibfnamefont {A.}~\bibnamefont
  {Gozolchiani}}, \bibinfo {author} {\bibfnamefont {K.}~\bibnamefont
  {Yamasaki}}, \bibinfo {author} {\bibfnamefont {O.}~\bibnamefont {Gazit}}, \
  and\ \bibinfo {author} {\bibfnamefont {S.}~\bibnamefont {Havlin}},\ }\href
  {\doibase 10.1209/0295-5075/83/28005} {\bibfield  {journal} {\bibinfo
  {journal} {Europhys. Lett.}\ }\textbf {\bibinfo {volume} {83}},\ \bibinfo
  {pages} {28005} (\bibinfo {year} {2008})}\BibitemShut {NoStop}%
\bibitem [{\citenamefont {Tsonis}\ \emph {et~al.}(2008)\citenamefont {Tsonis},
  \citenamefont {Swanson},\ and\ \citenamefont {Wang}}]{tsonisteleconnection}%
  \BibitemOpen
  \bibfield  {author} {\bibinfo {author} {\bibfnamefont {A.}~\bibnamefont
  {Tsonis}}, \bibinfo {author} {\bibfnamefont {K.}~\bibnamefont {Swanson}}, \
  and\ \bibinfo {author} {\bibfnamefont {G.}~\bibnamefont {Wang}},\ }\href
  {\doibase 10.1175/2007JCLI1907.1} {\bibfield  {journal} {\bibinfo  {journal}
  {J. Climate}\ }\textbf {\bibinfo {volume} {21}},\ \bibinfo {pages} {2990}
  (\bibinfo {year} {2008})}\BibitemShut {NoStop}%
\bibitem [{\citenamefont {Donges}\ \emph {et~al.}(2011)\citenamefont {Donges},
  \citenamefont {Schultz}, \citenamefont {Marwan}, \citenamefont {Zou},\ and\
  \citenamefont {Kurths}}]{dongeshladiny}%
  \BibitemOpen
  \bibfield  {author} {\bibinfo {author} {\bibfnamefont {J.~F.}\ \bibnamefont
  {Donges}}, \bibinfo {author} {\bibfnamefont {H.~C.}\ \bibnamefont {Schultz}},
  \bibinfo {author} {\bibfnamefont {N.}~\bibnamefont {Marwan}}, \bibinfo
  {author} {\bibfnamefont {Y.}~\bibnamefont {Zou}}, \ and\ \bibinfo {author}
  {\bibfnamefont {J.}~\bibnamefont {Kurths}},\ }\href {\doibase
  10.1140/epjb/e2011-10795-8} {\bibfield  {journal} {\bibinfo  {journal} {Eur.
  Phys. J. B}\ }\textbf {\bibinfo {volume} {84}},\ \bibinfo {pages} {635}
  (\bibinfo {year} {2011})}\BibitemShut {NoStop}%
\bibitem [{\citenamefont {Malik}\ \emph {et~al.}(2012)\citenamefont {Malik},
  \citenamefont {Bookhagen}, \citenamefont {Marwan},\ and\ \citenamefont
  {Kurths}}]{malikmarwan}%
  \BibitemOpen
  \bibfield  {author} {\bibinfo {author} {\bibfnamefont {N.}~\bibnamefont
  {Malik}}, \bibinfo {author} {\bibfnamefont {B.}~\bibnamefont {Bookhagen}},
  \bibinfo {author} {\bibfnamefont {N.}~\bibnamefont {Marwan}}, \ and\ \bibinfo
  {author} {\bibfnamefont {J.}~\bibnamefont {Kurths}},\ }\href {\doibase
  10.1007/s00382-011-1156-4} {\bibfield  {journal} {\bibinfo  {journal} {Clim.
  Dyn.}\ }\textbf {\bibinfo {volume} {39}},\ \bibinfo {pages} {971} (\bibinfo
  {year} {2012})}\BibitemShut {NoStop}%
\bibitem [{\citenamefont {Carpi}\ \emph {et~al.}(2012)\citenamefont {Carpi},
  \citenamefont {Saco}, \citenamefont {Rosso},\ and\ \citenamefont
  {Ravetti}}]{tropevolv}%
  \BibitemOpen
  \bibfield  {author} {\bibinfo {author} {\bibfnamefont {L.}~\bibnamefont
  {Carpi}}, \bibinfo {author} {\bibfnamefont {P.}~\bibnamefont {Saco}},
  \bibinfo {author} {\bibfnamefont {O.}~\bibnamefont {Rosso}}, \ and\ \bibinfo
  {author} {\bibfnamefont {M.}~\bibnamefont {Ravetti}},\ }\href {\doibase
  10.1140/epjb/e2012-30413-7} {\bibfield  {journal} {\bibinfo  {journal} {Eur.
  Phys. J. B}\ }\textbf {\bibinfo {volume} {85}},\ \bibinfo {pages} {1}
  (\bibinfo {year} {2012})}\BibitemShut {NoStop}%
\bibitem [{\citenamefont {Barreiro}\ \emph {et~al.}(2011)\citenamefont
  {Barreiro}, \citenamefont {Marti},\ and\ \citenamefont
  {Masoller}}]{latinochaos}%
  \BibitemOpen
  \bibfield  {author} {\bibinfo {author} {\bibfnamefont {M.}~\bibnamefont
  {Barreiro}}, \bibinfo {author} {\bibfnamefont {A.~C.}\ \bibnamefont {Marti}},
  \ and\ \bibinfo {author} {\bibfnamefont {C.}~\bibnamefont {Masoller}},\
  }\href {\doibase 10.1063/1.3545273} {\bibfield  {journal} {\bibinfo
  {journal} {Chaos}\ }\textbf {\bibinfo {volume} {21}},\ \bibinfo {pages}
  {013101} (\bibinfo {year} {2011})}\BibitemShut {NoStop}%
\bibitem [{\citenamefont {Deza}\ \emph {et~al.}(2013)\citenamefont {Deza},
  \citenamefont {Barreiro},\ and\ \citenamefont {Masoller}}]{latinoepj}%
  \BibitemOpen
  \bibfield  {author} {\bibinfo {author} {\bibfnamefont {J.}~\bibnamefont
  {Deza}}, \bibinfo {author} {\bibfnamefont {M.}~\bibnamefont {Barreiro}}, \
  and\ \bibinfo {author} {\bibfnamefont {C.}~\bibnamefont {Masoller}},\ }\href
  {\doibase 10.1140/epjst/e2013-01856-5} {\bibfield  {journal} {\bibinfo
  {journal} {Eur. Phys. J.-Spec. Top.}\ }\textbf {\bibinfo {volume} {222}},\
  \bibinfo {pages} {511} (\bibinfo {year} {2013})}\BibitemShut {NoStop}%
\bibitem [{\citenamefont {Yamasaki}\ \emph
  {et~al.}(2009{\natexlab{b}})\citenamefont {Yamasaki}, \citenamefont
  {Gozolchiani},\ and\ \citenamefont {Havlin}}]{yamaphasyn}%
  \BibitemOpen
  \bibfield  {author} {\bibinfo {author} {\bibfnamefont {K.}~\bibnamefont
  {Yamasaki}}, \bibinfo {author} {\bibfnamefont {A.}~\bibnamefont
  {Gozolchiani}}, \ and\ \bibinfo {author} {\bibfnamefont {S.}~\bibnamefont
  {Havlin}},\ }\href {\doibase 10.1143/PTPS.179.178} {\bibfield  {journal}
  {\bibinfo  {journal} {Prog. Theor. Phys. Suppl.}\ ,\ \bibinfo {pages} {178}}
  (\bibinfo {year} {2009}{\natexlab{b}})}\BibitemShut {NoStop}%
\bibitem [{\citenamefont {Sarachik}\ and\ \citenamefont
  {Cane}(2010)}]{sarachik2010nino}%
  \BibitemOpen
  \bibfield  {author} {\bibinfo {author} {\bibfnamefont {E.~S.}\ \bibnamefont
  {Sarachik}}\ and\ \bibinfo {author} {\bibfnamefont {M.~A.}\ \bibnamefont
  {Cane}},\ }\href@noop {} {\emph {\bibinfo {title} {The {E}l
  {N}i{\~n}o-southern oscillation phenomenon}}}\ (\bibinfo  {publisher}
  {Cambridge University Press},\ \bibinfo {year} {2010})\BibitemShut {NoStop}%
\bibitem [{\citenamefont {{Scholz}}(2010)}]{nodesim}%
  \BibitemOpen
  \bibfield  {author} {\bibinfo {author} {\bibfnamefont {M.}~\bibnamefont
  {{Scholz}}},\ }\href@noop {} {\bibfield  {journal} {\bibinfo  {journal}
  {ArXiv e-prints}\ } (\bibinfo {year} {2010})},\ \Eprint
  {http://arxiv.org/abs/1010.0803} {arXiv:1010.0803 [physics.soc-ph]}
  \BibitemShut {NoStop}%
\bibitem [{\citenamefont {Jiang}\ \emph {et~al.}(1995)\citenamefont {Jiang},
  \citenamefont {Neelin},\ and\ \citenamefont {Ghil}}]{ensoqqqbo}%
  \BibitemOpen
  \bibfield  {author} {\bibinfo {author} {\bibfnamefont {N.}~\bibnamefont
  {Jiang}}, \bibinfo {author} {\bibfnamefont {J.}~\bibnamefont {Neelin}}, \
  and\ \bibinfo {author} {\bibfnamefont {M.}~\bibnamefont {Ghil}},\ }\href
  {\doibase 10.1007/BF00223723} {\bibfield  {journal} {\bibinfo  {journal}
  {Clim. Dyn.}\ }\textbf {\bibinfo {volume} {12}},\ \bibinfo {pages} {101}
  (\bibinfo {year} {1995})}\BibitemShut {NoStop}%
\bibitem [{\citenamefont {Kondrashov}\ \emph {et~al.}(2005)\citenamefont
  {Kondrashov}, \citenamefont {Kravtsov}, \citenamefont {Robertson},\ and\
  \citenamefont {Ghil}}]{kondrashov2005hierarchy}%
  \BibitemOpen
  \bibfield  {author} {\bibinfo {author} {\bibfnamefont {D.}~\bibnamefont
  {Kondrashov}}, \bibinfo {author} {\bibfnamefont {S.}~\bibnamefont
  {Kravtsov}}, \bibinfo {author} {\bibfnamefont {A.~W.}\ \bibnamefont
  {Robertson}}, \ and\ \bibinfo {author} {\bibfnamefont {M.}~\bibnamefont
  {Ghil}},\ }\href {\doibase 10.1175/JCLI3567.1} {\bibfield  {journal}
  {\bibinfo  {journal} {J. Climate}\ }\textbf {\bibinfo {volume} {18}},\
  \bibinfo {pages} {4425} (\bibinfo {year} {2005})}\BibitemShut {NoStop}%
\bibitem [{\citenamefont {Sheppard}\ \emph {et~al.}(2011)\citenamefont
  {Sheppard}, \citenamefont {Stefanovska},\ and\ \citenamefont
  {McClintock}}]{harmonics}%
  \BibitemOpen
  \bibfield  {author} {\bibinfo {author} {\bibfnamefont {L.~W.}\ \bibnamefont
  {Sheppard}}, \bibinfo {author} {\bibfnamefont {A.}~\bibnamefont
  {Stefanovska}}, \ and\ \bibinfo {author} {\bibfnamefont {P.~V.~E.}\
  \bibnamefont {McClintock}},\ }\href {\doibase 10.1103/PhysRevE.83.016206}
  {\bibfield  {journal} {\bibinfo  {journal} {Phys. Rev. E}\ }\textbf {\bibinfo
  {volume} {83}},\ \bibinfo {pages} {016206} (\bibinfo {year}
  {2011})}\BibitemShut {NoStop}%
\bibitem [{\citenamefont {Palu\v{s}}\ and\ \citenamefont
  {Novotn\'a}(2007)}]{jastp1}%
  \BibitemOpen
  \bibfield  {author} {\bibinfo {author} {\bibfnamefont {M.}~\bibnamefont
  {Palu\v{s}}}\ and\ \bibinfo {author} {\bibfnamefont {D.}~\bibnamefont
  {Novotn\'a}},\ }\href {\doibase DOI 10.1016/j.jastp.2007.05.009} {\bibfield
  {journal} {\bibinfo  {journal} {J. Atmos. Sol.-Terr. Phys.}\ }\textbf
  {\bibinfo {volume} {69}},\ \bibinfo {pages} {2405} (\bibinfo {year}
  {2007})}\BibitemShut {NoStop}%
\bibitem [{\citenamefont {Marshall}\ \emph {et~al.}(2001)\citenamefont
  {Marshall}, \citenamefont {Kushnir}, \citenamefont {Battisti}, \citenamefont
  {Chang}, \citenamefont {Czaja}, \citenamefont {Dickson}, \citenamefont
  {Hurrell}, \citenamefont {McCartney}, \citenamefont {Saravanan},\ and\
  \citenamefont {Visbeck}}]{marshall2001}%
  \BibitemOpen
  \bibfield  {author} {\bibinfo {author} {\bibfnamefont {J.}~\bibnamefont
  {Marshall}}, \bibinfo {author} {\bibfnamefont {Y.}~\bibnamefont {Kushnir}},
  \bibinfo {author} {\bibfnamefont {D.}~\bibnamefont {Battisti}}, \bibinfo
  {author} {\bibfnamefont {P.}~\bibnamefont {Chang}}, \bibinfo {author}
  {\bibfnamefont {A.}~\bibnamefont {Czaja}}, \bibinfo {author} {\bibfnamefont
  {R.}~\bibnamefont {Dickson}}, \bibinfo {author} {\bibfnamefont
  {J.}~\bibnamefont {Hurrell}}, \bibinfo {author} {\bibfnamefont
  {M.}~\bibnamefont {McCartney}}, \bibinfo {author} {\bibfnamefont
  {R.}~\bibnamefont {Saravanan}}, \ and\ \bibinfo {author} {\bibfnamefont
  {M.}~\bibnamefont {Visbeck}},\ }\href {\doibase 10.1002/joc.693} {\bibfield
  {journal} {\bibinfo  {journal} {Int. J. Clim.}\ }\textbf {\bibinfo {volume}
  {21}},\ \bibinfo {pages} {1863} (\bibinfo {year} {2001})}\BibitemShut
  {NoStop}%
\bibitem [{\citenamefont {Fernandez}\ \emph {et~al.}(2003)\citenamefont
  {Fernandez}, \citenamefont {Hernandez},\ and\ \citenamefont
  {Pacheco}}]{naopink}%
  \BibitemOpen
  \bibfield  {author} {\bibinfo {author} {\bibfnamefont {I.}~\bibnamefont
  {Fernandez}}, \bibinfo {author} {\bibfnamefont {C.~N.}\ \bibnamefont
  {Hernandez}}, \ and\ \bibinfo {author} {\bibfnamefont {J.~M.}\ \bibnamefont
  {Pacheco}},\ }\href {\doibase
  http://dx.doi.org/10.1016/S0378-4371(03)00056-6} {\bibfield  {journal}
  {\bibinfo  {journal} {Physica A}\ }\textbf {\bibinfo {volume} {323}},\
  \bibinfo {pages} {705 } (\bibinfo {year} {2003})}\BibitemShut {NoStop}%
\bibitem [{\citenamefont {Feliks}\ \emph {et~al.}(2013)\citenamefont {Feliks},
  \citenamefont {Groth}, \citenamefont {Robertson},\ and\ \citenamefont
  {Ghil}}]{feliks13}%
  \BibitemOpen
  \bibfield  {author} {\bibinfo {author} {\bibfnamefont {Y.}~\bibnamefont
  {Feliks}}, \bibinfo {author} {\bibfnamefont {A.}~\bibnamefont {Groth}},
  \bibinfo {author} {\bibfnamefont {A.~W.}\ \bibnamefont {Robertson}}, \ and\
  \bibinfo {author} {\bibfnamefont {M.}~\bibnamefont {Ghil}},\ }\href {\doibase
  10.1175/JCLI-D-13-00105.1} {\bibfield  {journal} {\bibinfo  {journal} {J.
  Climate}\ }\textbf {\bibinfo {volume} {26}},\ \bibinfo {pages} {9528}
  (\bibinfo {year} {2013})}\BibitemShut {NoStop}%
\bibitem [{\citenamefont {Xu}\ \emph {et~al.}(2006)\citenamefont {Xu},
  \citenamefont {Chen}, \citenamefont {Hu}, \citenamefont {Stanley},\ and\
  \citenamefont {Ivanov}}]{PhysRevE.73.065201}%
  \BibitemOpen
  \bibfield  {author} {\bibinfo {author} {\bibfnamefont {L.}~\bibnamefont
  {Xu}}, \bibinfo {author} {\bibfnamefont {Z.}~\bibnamefont {Chen}}, \bibinfo
  {author} {\bibfnamefont {K.}~\bibnamefont {Hu}}, \bibinfo {author}
  {\bibfnamefont {H.~E.}\ \bibnamefont {Stanley}}, \ and\ \bibinfo {author}
  {\bibfnamefont {P.~C.}\ \bibnamefont {Ivanov}},\ }\href {\doibase
  10.1103/PhysRevE.73.065201} {\bibfield  {journal} {\bibinfo  {journal} {Phys.
  Rev. E}\ }\textbf {\bibinfo {volume} {73}},\ \bibinfo {pages} {065201}
  (\bibinfo {year} {2006})}\BibitemShut {NoStop}%
\bibitem [{\citenamefont {Martin}\ \emph {et~al.}(2013)\citenamefont {Martin},
  \citenamefont {Paczuski},\ and\ \citenamefont {Davidsen}}]{martindavidsen}%
  \BibitemOpen
  \bibfield  {author} {\bibinfo {author} {\bibfnamefont {E.~A.}\ \bibnamefont
  {Martin}}, \bibinfo {author} {\bibfnamefont {M.}~\bibnamefont {Paczuski}}, \
  and\ \bibinfo {author} {\bibfnamefont {J.}~\bibnamefont {Davidsen}},\ }\href
  {\doibase 10.1209/0295-5075/102/48003} {\bibfield  {journal} {\bibinfo
  {journal} {Europhys. Lett.}\ }\textbf {\bibinfo {volume} {102}},\ \bibinfo
  {pages} {48003} (\bibinfo {year} {2013})}\BibitemShut {NoStop}%
\end{thebibliography}

%

\end{document}